\documentclass[11pt,3p,times,authoryear]{elsarticle}

\usepackage{amssymb}
\usepackage{hyperref}
\usepackage{verbatim}
\usepackage{amsmath,esint}
\usepackage{yfonts}
\usepackage{units}
\usepackage{verbatim}
\usepackage{graphicx}
\usepackage{bm}
\usepackage{soul} % para tachar palabras
\usepackage{subcaption}
\usepackage{mathtools, cuted}
\usepackage{todonotes}
\usepackage{float}

\usepackage{wrapfig}
\usepackage{enumerate}
\usepackage{multirow}
\usepackage{booktabs}
\usepackage{array}
\usepackage{tabulary}
\usepackage{xspace}
\usepackage[normalem]{ulem}
\usepackage{lineno}
\usepackage{breakurl}
\usepackage{footnote}
\usepackage{tablefootnote}
\usepackage{threeparttable}
\usepackage{accents}

\newlength{\dhatheight}
\newcommand{\doublehat}[1]{%
    \settoheight{\dhatheight}{\ensuremath{\hat{#1}}}%
    \addtolength{\dhatheight}{-0.2ex}%
    \hat{\vphantom{\rule{1pt}{\dhatheight}}%
    \smash{\hat{#1}}}}

\newcommand{\doublehatnu}{\doublehat{\bm{\nu}}}
\newcommand{\hatnu}{\hat{\bm{\nu}}}

\newcommand{\data}{\bm{\mathcal{D}}}
\newcommand{\lkl}{\mathcal{L}}

\newcommand{\mdm}{m_\mathrm{DM}}
\newcommand{\tdm}{\tau_\mathrm{DM}}
\newcommand{\taudm}{\tau_\mathrm{DM}}

\newcommand{\aeff}{ A_\mathrm{eff}}

\newcommand{\bb}{b\bar{b}}
\newcommand{\WW}{W^+W^-}
\newcommand{\tautau}{\tau^+\tau^-}
\newcommand{\mumu}{\mu^+\mu^-}

\newcommand{\Tobsi}{T_{\mathrm{obs},i}}

\newcommand{\Nonij}{{N_{\mathrm{ON},ij}}}
\newcommand{\Noffij}{{N_{\mathrm{OFF},ij}}}
\newcommand{\Nbins}{N_\mathrm{bins}}
\newcommand{\gij}{g_{ij}}
\newcommand{\goffij}{g_{ij}^{\mathrm{OFF}}}
\newcommand{\bij}{b_{ij}}
\newcommand{\fij}{f_{ij}}
\newcommand{\tauobsi}{\kappa_{\mathrm{obs},i}}
\newcommand{\sigmataui}{\sigma_{\kappa,i}}
\newcommand{\sigmatausys}{\sigma_{\kappa}^{\mathrm{sys}}}

\newcommand{\epMinj}{{E'_{\mathrm{min},j}}}
\newcommand{\epMaxj}{{E'_{\mathrm{max},j}}}
\newcommand{\tauDM}{\tau_\mathrm{DM}}

\newcommand{\GeV}{\mbox{GeV}}
\newcommand{\TeV}{\mbox{TeV}}
\newcommand{\muG}{\mu\mbox{G}}

\let\olditemize\itemize
\renewcommand{\itemize}{
  \olditemize
  \setlength{\itemsep}{1pt}
  \setlength{\parskip}{0pt}
  \setlength{\parsep}{0pt}
}

\usepackage{lineno}

\usepackage[normalem]{ulem} % for strikethrough
\usepackage{adjustbox}
\usepackage{multirow}
\newcommand{\strike}{\bgroup\markoverwith{\textcolor{red}{\rule[0.5ex]{2pt}{0.4pt}}}\ULon}

\usepackage{color,xcolor}
\definecolor{amethyst}{rgb}{0.6, 0.4, 0.8}
\definecolor{green}{rgb}{0.55, 0.71, 0.0}
\definecolor{apricot}{rgb}{0.98, 0.81, 0.69}
\definecolor{auburn}{rgb}{0.43, 0.21, 0.1}
\definecolor{babyblueeyes}{rgb}{0.63, 0.79, 0.95}
\definecolor{bittersweet}{rgb}{1.0, 0.44, 0.37}
\journal{Physics of the Dark Universe}

\begin{document}

\begin{frontmatter}
\title{Constraining Dark Matter lifetime with a deep gamma-ray survey of the Perseus Galaxy Cluster with MAGIC}

%\include{authors}
% authors 22.01.2018  Format AA
%

%\author{M.~Doro, J.~Palacio, J.~Rico, M.~Vazquez Acosta}

\author{MAGIC Collaboration:}
\author[a]{V.~A.~Acciari}
\author[b,u]{S.~Ansoldi}
\author[c]{L.~A.~Antonelli}
\author[d]{A.~Arbet Engels}
\author[e]{C.~Arcaro}
\author[f]{D.~Baack}
\author[g]{A.~Babi\'c}
\author[h]{B.~Banerjee}
\author[i]{P.~Bangale}
% ...
%}
\address[a]{Inst. de Astrof\'isica de Canarias, E-38200 La Laguna, and Universidad de La Laguna, Dpto. Astrof\'isica, E-38206 La Laguna, Tenerife, Spain} 
\address[b]{Universit\`a di Udine, and INFN Trieste, I-33100 Udine, Italy} 
\address[c]{National Institute for Astrophysics (INAF), I-00136 Rome, Italy}
\address[d]{ETH Zurich, CH-8093 Zurich, Switzerland}
\address[e]{Universit\`a di Padova and INFN, I-35131 Padova, Italy}
\address[f]{Technische Universit\"at Dortmund, D-44221 Dortmund, Germany}
\address[g]{Croatian MAGIC Consortium: University of Rijeka, 51000 Rijeka, University of Split - FESB, 21000 Split,  University of Zagreb - FER, 10000 Zagreb, University of Osijek, 31000 Osijek and Rudjer Boskovic Institute, 10000 Zagreb, Croatia.}
\address[h]{Saha Institute of Nuclear Physics, HBNI, 1/AF Bidhannagar, Salt Lake, Sector-1, Kolkata 700064, India}
\address[i]{Max-Planck-Institut f\"ur Physik, D-80805 M\"unchen, Germany}
% ...
\address[u]{Japanese MAGIC Consortium: ICRR, The University of Tokyo, 277-8582 Chiba, Japan; Department of Physics, Kyoto University, 606-8502 Kyoto, Japan; Tokai University, 259-1292 Kanagawa, Japan; RIKEN, 351-0198 Saitama, Japan}
% ...

\cortext[corr]{Corresponding authors: jpalacio@ifae.es, michele.doro@unipd.it, monicava@iac.es, jrico@ifae.es}

\begin{abstract}
Clusters of galaxies are the largest known gravitationally bound structures
in the Universe, with masses around $10^{15}$~M$_\odot$, most of it in the form of dark matter.
The ground-based Imaging Atmospheric Cherenkov Telescope MAGIC made a deep survey of the Perseus cluster
of galaxies using almost 400~h of data recorded between 2009 and 2017.
%We search for decaying dark matter
%using data from the ground-based Imaging Atmospheric Cherenkov Telescopes MAGIC 
%recorded between 2009 and 2017,
%we made a survey of the Perseus cluster of galaxies. 
This is 
the deepest observational campaign so far on a cluster of galaxies in the very high energy range.
%This is the deepest campaign carried on a cluster of galaxies performed so far in the very high 
%energy ($E>50$~GeV) regime from any ground-based gamma-ray instruments. 
We search for gamma-ray signals from dark matter particles in the mass range between 
200~GeV and 200~TeV decaying into standard model pairs. 
We apply an analysis optimized for the spectral and morphological features expected 
from dark matter decays and find no evidence of decaying dark matter.
From this, we conclude that dark matter particles have a decay lifetime longer than $\sim10^{26}$~s in all considered channels.
Our results improve previous lower limits found by MAGIC 
and represent the strongest limits on decaying dark matter particles
from ground-based gamma-ray instruments.
\end{abstract}

\begin{keyword}
decaying dark matter
\sep 
cluster of galaxies
\sep
indirect searches
\sep
Imaging Air Cherenkov Telescopes
\sep
Perseus
%% keywords here, in the form: keyword \sep keyword
%% PACS codes here, in the form: \PACS code \sep code
%% MSC codes here, in the form: \MSC code \sep code
%% or \MSC[2008] code \sep code (2000 is the default)
\end{keyword}

\end{frontmatter}

%\linenumbers
%\input{comments}
\section{Introduction}
\label{sec:introduction}
Decades of observational evidence show that the Standard Model (SM) of Particles Physics cannot entirely explain 
the gravitational balance observed at all cosmological scales, from that of Milky Way satellite dwarf spheroidal galaxies (dSphs) 
to that of cluster of galaxies ~\citep[CGs, see][]{Roos:2010,Freese:2008cz}.
In order to explain these observations, Dark Matter (DM) has been suggested to exist in the form of a new elementary particle,
currently only seen through its gravitational imprint. 
Weakly-Interacting Massive Particles (WIMPs) are generic massive particles with an expected mass range between few GeV \citep[Lee-Winberg limit,
see][]{Boehm:2002yz} and few hundreds of TeV \citep[unitary bound, see][]{Griest:1989wd}. 
WIMPs are expected to interact with SM particles with strengths at the weak scale,
and to be either stable or very long lived. 
A WIMP 
can either annihilate or decay into SM particles, or even be decoupled from the SM.
%totally secluded in a dark sector 
%\textbf{(totally isolated from interacting with SM %particles)}. 
The WIMP paradigm has been long debated, as the WIMP self-annihilation in the early Universe naturally accounts
for the DM density observed at present (typically referred to as the \emph{WIMP miracle}), 
%at the same time 
being possibly within reach of different currently operating instruments.
The case of DM annihilation has received greater attention in the literature~\citep{Feng:2010gw}
but there is no experimental or theoretical guarantee that DM particles are absolutely stable.  
%Therefore, decaying DM signals still provide a valuable experimental window~\citep{Ibarra:2012a}.
The only constraint is that decaying DM particles' lifetime should be comparable or larger than the Hubble 
time of $\sim10^{17}$~s in order to explain the current DM density.
%Still, decaying DM could leave a trace or particles detectable from Earth. 
Among others, decaying DM particles may produce e.g. leptons, quarks, or gauge bosons, 
which can subsequently provide electromagnetic radiation due to prompt emission or secondary interactions. %~\citep{Ibarra:2013cra}. 
Lately, DM models that favour decays into leptons (known as ``leptophilic'' models) have received increased 
attention, due to the excess of positron events observed in the local cosmic ray (CR) flux by PAMELA, AMS-II and
Fermi-LAT~\citep{Adriani:2008zr,Abdo:2009zk,Ackermann:2010ij,Aguilar:2013qda}. 
\newline

The standard cosmological model 
%Hierarchical structure formation 
predicts CGs to be 
the latest and most massive structures to form in the Universe \citep{Peebles:1994xt}.
With higher DM concentration and closer distances, dSphs and the Galactic Center are among the best regions to search for annihilating WIMPs.
CGs however, with masses of the order of $10^{14-15}$~M$_\odot$~\citep[$\sim$80$\%$ of it in the form of DM, see e.g.,][]{Jeltema:2008vu,Pinzke:2009cp}, 
are excellent laboratories to study decaying DM. 
The Perseus CG is a cool-core cluster located at a distance of 77.7~Mpc (redshift $z=0.0183$). 
Perseus is very bright in X-rays, and one of the best candidates for
detecting CR induced gamma rays that come from particle acceleration at the cluster core~\citep{Aleksic:2009ir,Pinzke:2010st,Pinzke:2011ek}.
The Perseus CG is considered among the most promising CGs for gamma-ray
indirect DM detection~\citep{SanchezConde:2011ap}.
\newline

The Major Atmospheric Gamma Imaging Cherenkov (MAGIC) telescopes~\citep{Aleksic:2014lkm}
%\footnote{\href{https://magic.mpp.mpg.de/}{https://magic.mpp.mpg.de/}} 
have observed the Perseus CG 
since 2009, the deepest exposure the instrument has carried out.
The campaign took place over several consecutive years and comprised almost 400~h of recorded data %(see~\autoref{tab:data}) 
until 2017.
%The MAGIC telescopes carried out a long observation campaign on the Perseus CG between 2009 and 2017,
%collecting nearly $400$~h data.
MAGIC is a system of two 17~m diameter Imaging Atmospheric Cherenkov Telescopes (IACTs) capable of detecting gamma rays in the very high 
energy (VHE, E$>50$~GeV) band. % interacting with the Earth's atmosphere.
For low zenith angle observations, MAGIC has 
an angular resolution of $\sim$0.1$^\circ$,
a trigger threshold of $\sim$50~GeV,
and sensitivity for point-like sources %with Crab Nebula-like spectrum 
of $\sim$0.66\% of Crab Nebula flux above 220~GeV in 50~h of observation~\citep{Aleksic:2014lkm}.
The MAGIC campaign on Perseus CG 
proved to be very fruitful, producing the strongest limits on CR 
acceleration and CR pressure in the core of the cluster~\citep{Aleksic:2009ir,Aleksic:2011cp};
a clear detection and model for the radio galaxy NGC~1275, at the center of the 
cluster~\citep{Aleksic:2011eb, Aleksic:2013kaa};
and the detection of the peculiar radio galaxy IC~310, located at 0.6~deg from the Perseus CG 
center, which provides important evidence related to the acceleration of CRs close to black 
holes~\citep{Aleksic:2010xk, Aleksic:2013bya, Aleksic:2014xsg}.
\newline

%In this work we focus on the signatures of decaying DM from the Perseus CG.  
%DM particles may either decay into SM pairs ($b\bar{b}$, $\tau^+\tau^-$, $\mu^+\mu^-$,
%$W^+W^-$), producing rather continuous spectra of gamma rays reaching the Earth,
%or produce a line-like spectra emerging from decays into $\gamma\gamma$, $\gamma\nu$ or $Z\gamma$.
In this work we focus on the search for signatures of decaying DM in an extended region from the Perseus CG
with observations from MAGIC.  
We do not consider the annihilation case since 
the expected signal of DM annihilation in the Perseus CG would be one order of 
magnitude smaller than the signal expected for the typical case of dSphs~\citep{SanchezConde:2011ap}. 
More importantly, the expected morphology of the signal of annihilating DM is more concentrated 
towards the center of the cluster that of decay where, 
in the case of the Perseus CG, we have limited sensitivity due to the presence of the variable flux 
gamma-ray emitter NGC~1275. %, located at the center of the CG.
We search for signatures of decaying DM 
particles in the mass range between $200$~GeV and $200$~TeV for hadronic/leptonic decays,
and for DM particles with masses between $200$~GeV and $20$~TeV decaying into $\gamma\gamma$.
We find no evidence for a DM signal and consequently set 95\% confidence level (CL) lower-limits on the decay lifetime 
of the DM particle for the different assumed mass values and decay channels.

%
% PAPER STRUCTURE
%
\bigskip

\noindent
The rest of this paper is structured as follows. 
\autoref{sec:dm} discusses the expected DM distribution of the Perseus CG and the photon flux at 
Earth coming from DM decays.
\autoref{sec:data} describes the MAGIC observations and event selection, optimized to search for DM decays.
In \autoref{sec:BinnedLikelihoodAnalysis} we introduce the likelihood formalism used in the analysis.
In \autoref{sec:results} we present the obtained 
lower limits on the decay lifetime, which are put into context with other measurements.
In \autoref{sec:conclusions} the paper is briefly summarized and the conclusions are given.

\section{Expected Dark Matter Decay Signal from Perseus}
\label{sec:dm}
\noindent
In order to compute the gamma-ray rate expectations and tailor the data reconstruction and analysis, we
first need to assess the DM framework.
The differential gamma-ray flux coming from decaying DM from a given direction in the sky 
%of the Perseus CG 
is given by:
\begin{eqnarray}\label{eq:diffflux}
\nonumber
\frac{\text{d}^{2}\Phi}{\text{d}E \, \text{d}\Omega}= \frac{1}{4\pi}\, \frac{1}{\taudm\; \mdm}\; 
\frac{\mathrm{d}N_\gamma}{\mathrm{d}E} \frac{\mathrm{d}J_{\mbox{dec}}}{\mathrm{d}\Omega}  ,\\
\end{eqnarray}
%
%\begin{equation}\label{eq:flux}
%\frac{d\Phi_\gamma}{dE}(E,\Delta\Omega)=\ppterm \cdot J(\Delta\Omega), 
%\end{equation}
%
\noindent
%where $\sppterm$ is called
%the \emph{particle-physics factor}, which gives the DM decay rate and depends on the nature of the DM particle,
where $\mdm$ is the DM mass, $\taudm$ the DM particle lifetime,
${\text{d}N_{\gamma}}/{\text{d}E}$ is the average decay spectrum per reaction,
and ${\text{d}J_{\mbox{dec}}}/{\text{d}\Omega}$ is called the \emph{differential astrophysical 
factor}~\citep[or simply differential $J$-factor,][]{Bergstrom:1997fj} and
is obtained integrating the DM density $\rho$ over the line-of-sight (l.o.s.) for the decay reaction:
%The DM particle identity in~\autoref{eq:diffflux} is contained in $\mdm$, $\taudm$, and ${\text{d}N_{\gamma}}/{\text{d}E}$, 
%whereas ${\text{d}J_{\mbox{dec}}}/{\text{d}\Omega}$ depends on the distribution of DM and its distance to Earth as
\begin{eqnarray}\label{eq:PPandJfactor}
\frac{\text{d}J_{\mbox{dec}}}{\text{d}\Omega} & =&   \int_\mathrm{l.o.s.}  \,\mathrm{d}l\,
  \rho(l,\Omega).
\end{eqnarray}
The total $J$-factor enclosed in a given sky region 
can be obtained integrating \autoref{eq:PPandJfactor}
%over the line-of-sight (l.o.s.) and 
over a solid angle $\Delta\Omega$.
%The total $J$-factor %we refer is computed within a defined region in the sky.
%This quantity however, 
%is irrelevant for the analysis, since this has to be necessarily computed in reconstructed distance, which we later do.
%Qualitatively, an effective $J$-factor ($J_{\text{eff}}$) should be computed taking into account the instrument point spread function, 
%but this also depends on the energy and hence, $J_{\text{eff}}$ would depend on each WIMP decay spectra and mass.
%Because of that, for extended sources higher $J$-factor does not necessarily imply higher sensitivity.
%
%\begin{eqnarray}\label{eq:PPandJfactor}
%J_{dec}(\dom) & =&  \int_{\dom}  \ d\Omega \ \int_\mathrm{l.o.s.}  \,dl\,
%  \rho(l,\Omega).
%\end{eqnarray}
We proceed now in discussing the terms of \autoref{eq:diffflux} separately. 
%
%
%
%
%
%\noindent
\subsection{Decay DM particle models}
\label{sec:dm_models}
%\textbf{(valuable models for decaying DM are X, Y, Z see .e.g \cite{x,y,z})}
%As mentioned before, 
%The microscopic nature of the DM particle  in~\autoref{eq:diffflux} is contained in $\mdm$, $\taudm$, and ${\text{d}N_{\gamma}}/{\text{d}E}$. 
An important constraint for decaying DM comes from the fact that, if 
DM particles were in thermal equilibrium in the early Universe, 
$\taudm$ should be larger than the age of the Universe in order to explain the current 
observed DM density. 
Scenarios with such candidates should only allow for a small violation of their stability~\citep{Berezinsky:1991sp,Chen:2003gz,Ando:2015qda}. 
Alternatively, the stability can be related to the strength of the space-time
curvature enabling the so-called gravity portal~\citep{Cata:2016epa}. 
%\textbf{(explain gravity portal¿?)} 
There are several valid candidates for decaying DM proposed in the literature~\citep[see, e.g.,][]{Feng:2010gw}, such as the sterile neutrino, the axion, or the 
super-symmetric candidates gravitino, lightest right-handed sneutrino, and wino.
Regardless the microscopic nature of the particle,
%
%Generally speaking,  
%continuous gamma-ray spectra from 
DM decays can be roughly classified as 
$(i)$ leptonic, $(ii)$ hadronic, or $(iii)$ a mix of the two, 
according to the particle predominance in the decay products. 
DM candidates with masses 
at the TeV scale and leptophilic decay modes have been proposed
in order to match the measured spectral features seen in CR positron data~\citep[see e.g.][]{Feng:2014zca}.

%\begin{comment}
%Among the studied extensions of the SM theory is the so-called Supersymmetric SM %(SUSY), a theory that links gravity with 
%the other fundamental forces of nature postulating the existence of a new %\emph{super}-partner for each SM particle. 
%Typically, the Minimal SUSY extension (MSSM) is considered, since the number of free %(known) parameters is largely reduced w.r.t. the
%full theory.
%In the MSSM, baryon number and lepton number are no longer conserved once loop %corrections are taken into account,
%where, in order not to be in conflict with experimental data, a new discrete symmetry
%($R$-parity), acting on the fields that forbids these couplings, is also defined.
%With $R$-parity being preserved, the Lightest Supersymmetric Particle (LSP) cannot %decay, providing a viable DM candidate,
%typically the Neutralino.
%$R$-parity violation provides with DDM, Neutralino, or other possible LSP candidates %are the gravitino, the sterile/right-handed sneutrino or the axino depending on the %parameters of the model [ref].
%\end{comment}
\bigskip

Additionally, 
%Two-body decay processes involving photons often 
decay processes may also give rise to monochromatic photon lines~\citep{Garny:2011}.
%showed that leptophilic decaying DM can produce a potentially observable gamma-ray line signal.
Among the candidates previously mentioned, %that may produce such lines are 
the sterile-neutrino~\citep{Ando:2010ye} or the gravitino with masses below $2$-$300$ GeV~\citep{Ibarra:2012a}
could produce such a clear signal
%Gravitino candidates with masses below a $2-300$ GeV, show additionally, 
%a prominent decay into a gamma-ray line~\citep{Ibarra:2012a}.
%More generally,  
%The observation of lines would provide compelling evidence for DM  because 
that hardly any astrophysical process can mimic. 
\subsection{Dark Matter Distribution}
\label{sec:dm_distribution}
The DM density profile of different astrophysical sources (e.g., the Galactic Center, dSphs
%, globular clusters 
and CGs) 
is generally considered universal and can be expressed by a Zhao-Hernquist functional
form~\citep{Hernquist:1990be,Zhao:1995cp} 
%\textbf{(was this proposed for DM? check!)\md{yes}}
as
% There is a general agreement from N-body simulations, yet only partially supported with experimental data, that the DM density
% profile around different classes of DM dominated targets (from dwarf galaxies,
% to galaxies, and to clusters of galaxies) is universal, and can be
% expressed by the Zhao-Hernquist functional form~\citep[Zhao-Hern, see][]{Hernquist:1990be,Zhao:1995cp}.m
%called the Navarro-Frenk-White (NFW,
%\citep{Navarro:1995iw}) profile, which reads as:
%
\begin{equation}\label{eq:zhao}
\rho(r)=\frac{\rho_0}{\left(\frac{r}{r_s}\right)^\gamma\;
\left[ 1+\left(\frac{r}{r_s}\right)^\alpha\right]^{(\beta-\gamma)/\alpha}},
\end{equation}
% \begin{equation}\label{eq:nfw}
%   \rho(r)=\rho_{-2}\exp{\left(-\frac{2}{\alpha}\left[\left(\frac{r}{r_{-2}}\right)^\alpha-1\right]\right)},
%  \rho(r)=\frac{\rho_0}{r/r_s\;(1+r/r_s)^2}
% \end{equation}
%
where $r$ is the distance from the DM dynamical center of the cluster, 
$r_{s}$ and $\rho_{0}$ are the characteristic
scale radius and DM density, and $\alpha,\beta,\gamma$ are free parameters. %~\citep{Kuhlen:2012ft}.
%that take the value of $1,3,1$ for the often-used . Navarro-Frenk-White (NFW, \citep{Navarro:1995iw}) profile. 
%\jp{This assumption is driven by N-body simulations~\citep{Kuhlen:2012ft} and yet requires experimental evidence.}
%~\citep{X}% LSS -> Percival:2001hw,Cole:2005sx}.
Due to hierarchical structure formation, the total DM profile expressed in \autoref{eq:PPandJfactor} is the sum of 
a smooth component and a second component due to a large expected number of small DM substructures. 
The effect of DM substructures in the case of DM annihilation in CGs can increase the total astrophysical factor for annihilation $J_{\text{ann}}$ 
by a factor up to a few tens~\citep{SanchezConde:2011ap,Moline:2016pbm}.
% Pinzke:2011,Gao:2011rf,
In the case of decaying DM however, because of the linear dependence with the DM density (see~\autoref{eq:PPandJfactor}), substructures tend to 
average out for large observation angles and do not have a sizeable effect on $J_{\text{dec}}$. 
\newline

%For the computation of the astrophysical factor,
We follow the prescription in~\cite{SanchezConde:2011ap}, 
where the DM density profile of the Perseus
CG is modelled with a Navarro-Frenk-White parametrization~\citep[e.g. a Zhao-Hernquist profile with $\alpha=1$, $\beta=3$ and $\gamma=1$, ][]{Navarro:1995iw} 
with $r_s=0.477~\mbox{Mpc}$ and $\rho_0=7.25\times10^{14}~\mbox{M$_\odot\,$Mpc$^{-3}$}$. 
In our analysis we consider the entire DM halo of the Perseus cluster (with a radius of $\sim1.5^\circ$), 
which results in total decay $J$-factor of $1.5\times10^{19}~\mbox{GeV}\,\mbox{cm}^{-2}$. 
During the analysis, further angular cuts are applied, which will effectively reduce that value (see~\autoref{subsec:eventSelecion}).
%\textbf{The integration over the entire DM halo of the Perseus CG (over a radius of an angle of $\sim1.5^\circ$) results in a total decay $J$-factor
%\footnote{
%It should be clarified that 
%The total $J$-factor we refer is computed within a defined region in the sky.
%This quantity however, is irrelevant for the analysis, since this has to be necessarily computed in reconstructed distance, which we later do.
%Instead, an effective $J$-factor ($J_{\text{eff}}$) should be computed taking into account the instrument point spread function, 
%but this also depends on the energy and hence, $J_{\text{eff}}$ would depend on each WIMP decay spectra and mass.
%Because of that, for extended sources higher $J$-factor does not necessary imply higher sensitivity.}
%\footnote{
%\textbf{
%The total $J$-factor is only reported to provide a reference, but it is not directly used for computation of the upper limits. 
%In the likelihood an effective J-factor is used, that includes the instrument angular resolution, energy resolution as well as the DM decay spectrum.
%}
%}.}
\newline

The estimation of $J_{\text{dec}}$ is proportional to the total DM
mass in the source, and hence this is the largest source of uncertainty.
%Uncertainties in the density profile propagate linear to the $J_{\text{dec}}$-factor. 
In order to be considered in our analysis,
these uncertainty should be known as a function of the integration angle ($\Delta\Omega$ in~\autoref{eq:PPandJfactor}),
which is not the case.
Mass estimates for CGs show agreement of the order of 4\% uncertainty between lensing and hydrostatic estimation on a sample of 50 CGs~\citep{Smith:2015qhs}. 
However, the Perseus CG was not included in this study 
likely due to its vicinity or ample extension in the sky. 
For this reason, current available measurements of the total mass of the Perseus CG~\citep{Reiprich:1999qm,Chen:2007sz} 
have larger associated uncertainties of about 30\%.
%Current available measurements of the total mass of the Perseus CG~\citep{Reiprich:1999qm,Chen:2007sz}
%have associated uncertainties of about 30\% (considering systematic effects).
%As it will be discussed later, we cannot search for DM signal in the whole cluster extension. 
%In order to correctly propagate such uncertainties through the likelihood,
%Since we do not integrate the whole cluster extension, 
%in order to correctly propagate such uncertainties through the likelihood,
%they should be known as a function of the integration angle ($\Delta\Omega$ in~\autoref{eq:PPandJfactor}),
%which is not the case.
No uncertainties in the $J$-factor are considered in our analysis, but 
even assuming 
%We estimate that, in a conservative scenario, of 
a 50\% uncertainty on $J_{\text{dec}}$, 
our lower limits on the lifetime would be weakened by only a factor~2.

\section{MAGIC observation and data selection}
%\section{MAGIC observation and event selection}
\label{sec:data}
\begin{table*}[t]
  \footnotesize
  \centering
  % %\begin{tabular}{cc|ccc|ccc}
% %     \hline
% % %    &  & \multicolumn{2}{c|}{NGC-1275 ($\mathcal{A}$)} & \multicolumn{2}{c}{PERSEUS-MA ($\mathcal{B}$)}\\
% %     \multicolumn{2}{c|}{Perseus data sample} & \multicolumn{3}{c|}{$\mathcal{A}$} & \multicolumn{3}{c}{$\mathcal{B}$}\\
% %      \multicolumn{2}{c|}{} & All data [h] & \multicolumn{2}{c|}{Selected after [h]} & All data [h] & \multicolumn{2}{c}{Selected after [h]}\\
% %     IRF	& Period &  & NSB &  NSB+AOD &  & NSB &  NSB+AOD \\
% %     \hline
% %     ST.01.02 & 2009.11.01-2009.06.01 & 94.7 & 56.4 & 45.4 &  -  & - &  -    \\
% %     %ST.02.01 & 2012.01.19-2012.02.25 &  -   &  -   &  -   &  -   \\
% %     ST.03.01 & 2012.09.01-2013.01.17 & 9.2 & 9.1 &  9.1 & 59.4 & 40.2 & 36.8  \\
% %     %ST.03.02 & 2012.01.18-2012.07.26 &  -   &  -   &  -   &  -   \\
% %     ST.03.03 & 2013.07.27-2014.08.05 & 17.5 & 16.7 & 14.8 &  55 & 30.2 &  28.9  \\	
% %     ST.03.05 & 2014.08.31-2014.11.22 & 16.6 & 10.4 & 10.1 & 21.7 & 21.7  &  7.5  \\
% %     ST.03.06 & 2014.11.24-2016.04.28 & 6.8 & 3.9 &  3.9  & 29.3 & 22.32 & 21.9 \\ % (NSB+AOD) 4.7-> 3.9, check
% %     ST.03.07 & 2016.04.29-2017.08.02 & 44.1 & 41.9 & 12.2 & 20.5 & 16.02 & 11.1  \\ 
% %   	\hline
% %     TOTAL &	                   & 185.9 & 138.4 &  106.1 & 188.9 & 119.2 & 96.2 \\
% %     \hline
% %     \hline
% %     &\multicolumn{7}{c}{Global sample selected 202.2 h}\\
% % \hline
% %\end{tabular}
  \begin{tabular}{|c|c|c|c|c||c|c|c|}
    \cline{3-8}
    \multicolumn{2}{c}{} & \multicolumn{6}{c}{Telescope Pointing} \\
    \cline{3-8}
    \multicolumn{2}{c|}{} & \multicolumn{3}{c||}{$\mathcal{A}$} & \multicolumn{3}{c|}{$\mathcal{B}$}\\
    \hline
    & &  & \multicolumn{2}{|c||}{Data Selection} &  & \multicolumn{2}{|c|}{Data Selection}\\
    \cline{4-5} \cline{7-8}
    Period	& Dates & All data [h] & \multirow{2}{*}{quality [h]} & quality + & All data [h] &  \multirow{2}{*}{quality [h]}  &  quality + \\
    	&  &  &  &  specific [h] &  &  &  specific [h] \\
    \hline
    $\mathcal{P}1$ & 2009.11.01-2011.06.01 & 94.7 & 56.4 & 45.4 &  -  & - & - \\
    %ST.02.01 & 2012.01.19-2012.02.25 &  -   &  -   &  -   &  -   \\
    $\mathcal{P}2$ & 2012.09.01-2013.01.17 & 9.2 & 9.1 &  9.1 & 59.4 & 40.2 &  36.8 \\
    %ST.03.02 & 2012.01.18-2012.07.26 &  -   &  -   &  -   &  -   \\
    $\mathcal{P}3$ & 2013.07.27-2014.08.05 & 17.5 & 16.7 & 14.8 &  55 & 30.2 &   28.9 \\	
    $\mathcal{P}4$ & 2014.08.31-2014.11.22 & 16.6 & 10.4 & 10.1 & 21.7 & 21.7  &   7.5 \\
    $\mathcal{P}5$ & 2014.11.24-2016.04.28 & 6.8 & 3.9 &  3.9  & 29.3 & 22.32 & 21.9 \\ % (NSB+AOD) 4.7-> 3.9, check
    $\mathcal{P}6$ & 2016.04.29-2017.08.02 & 44.1 & 41.9 & 12.2 & 20.5 & 16.02 &  11.1 \\ 
   	\hline
    \multicolumn{2}{|c|}{TOTAL}           & 185.9 & 138.4 &  106.1 & 188.9 & 119.2 &  \multicolumn{1}{c|}{96.2}\\
    \hline
    \multicolumn{8}{c}{}\\
    \hline
    \hline
    \multicolumn{2}{c}{}&\multicolumn{2}{r}{Total selected}& \multicolumn{1}{l}{202.2 h}&\multicolumn{3}{c}{}\\
\hline
  \end{tabular}
\caption{\label{tab:data} Observations of the Perseus cluster with the MAGIC telescopes for two different telescope pointings $\mathcal{A}$ and $\mathcal{B}$ for different observational periods. 
The number of hours taken for each period and after data selection: \emph{quality} cuts are based on NSB and aerosol extinction,
\emph{specific} cuts are based on the night-wise significance of \emph{NGC1275}, \emph{NGC1265} or \emph{IC310} 
(see text for details).}
\end{table*}
\noindent

\begin{figure}[ht!]
  \centering
  \includegraphics[width=0.9\linewidth]{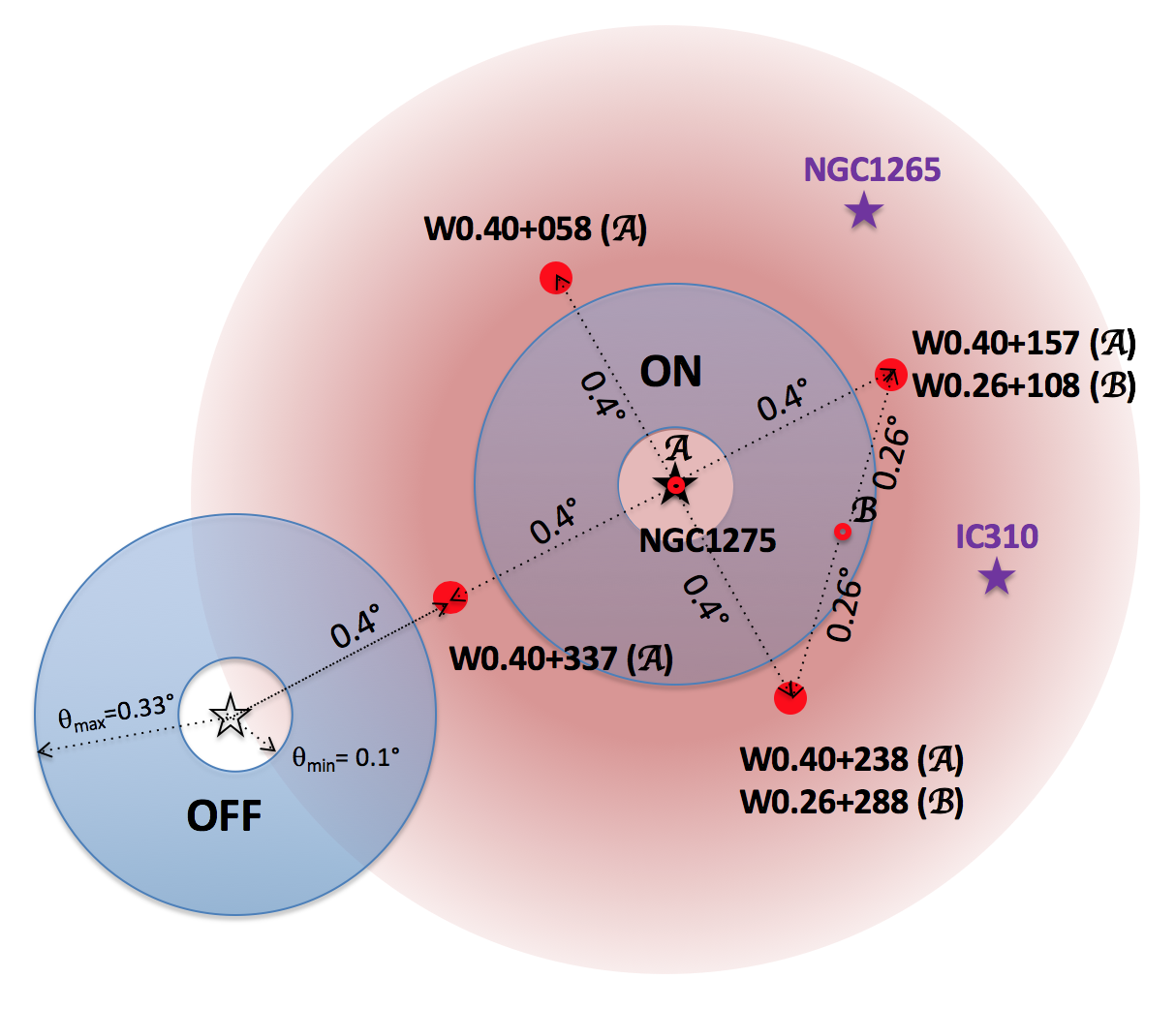}
    \caption{\label{fig:perseusFOV} 
    Schematic view of the Perseus CG FoV. 
    The location of the galaxies NCG~1275, IC~310, and NCG~1265 are marked with colored stars
    (the location of NCG~1275 is coincident with the center of the Perseus CG).
    The large blurred red region represents the expected DM decay signal morphology~\cite[based on][]{SanchezConde:2011ap}. 
    %The different pointing positions of the telescopes labelled \texttt{W0.40+XXX} (\texttt{W0.26+YYY}) for the two different pointing modes $\mathcal{A}$ ($\mathcal{B}$) are shown as red wide dots. 
    The nominal position of the two pointing modes labeled $\mathcal{A}$  and $\mathcal{B}$ are shown as open red circles whereas 
    the different pointing positions of the telescopes around these two pointing mode directions are labeled \texttt{W0.40+XXX} and \texttt{W0.26+YYY} 
    (for  pointing mode $\mathcal{A}$ and $\mathcal{B}$ respectively) and are shown as red wide dots.    
    ON/OFF regions from opposing pointings (e.g. ON from \texttt{W0.40+157} and OFF from \texttt{W0.40+337}, 
    where OFF center position is marked with an empty star) are analyzed in pairs.
    $\mathcal{R}$1 is the region around NCG~1275 defined by $\theta<\theta_{\mathrm{min}}$ (shown with dashed black arrows only for OFF) 
    with respect to NGC~1275's direction. %(which is coincident with the center of the cluster). 
    $\mathcal{R}$2 is the region defined between $\theta_{\mathrm{min}}<\theta<\theta_{\mathrm{max}}$ and are shown as blue regions for ON and OFF.
    %the two angular cuts $\theta_{\mathrm{min}}=0.1^{\circ}$ and $\theta_{\mathrm{max}}=0.33^{\circ}$. 
    Dark matter is searched within $\mathcal{R}$2 while $\mathcal{R}$1 is used to evaluate the gamma-ray emission activity of NGC~1275 for each given dataset.
    }
    % Schematic view of the Perseus CG $\FoV$.
    % The location of the galaxies NCG~1275, IC~310, and NCG~1265 are marked with yellow stars
    % (the location of NCG~1275 is coincident with the center of the Perseus CG).
    % The large green region represents the expected DM decay signal morphology~\cite[based on][]{SanchezConde:2011ap}. 
    % The different pointing positions of the telescopes labelled \texttt{W0.40+XXX} (\texttt{W0.26+YYY}) for pointing mode $\mathcal{A}$ ($\mathcal{B}$) are shown as red dots.
    % %Wobbles are analysed in pairs where, for the particular case of pointing \texttt{W0.40+337} ($\mathcal{A}$),
    % %ON and OFF regions are shown as a blue rings ($\mathcal{R}$2),
    % %defined by two angular cuts, $\theta_{\mathrm{min}}$ and $\theta_{\mathrm{max}}$ (shown with dashed black arrows only for OFF).
    % ON/OFF regions from opposing pointings (e.g. ON from \texttt{W0.40+157} and OFF from \texttt{W0.40+337},
    % as shown in the figure as blue regions, labelled $\mathcal{R}$2 in the OFF region)
    % are analyzed in pairs.
    % $\mathcal{R}$2 rings are defined by two angular cuts, $\theta_{\mathrm{min}}$ and $\theta_{\mathrm{max}}$ 
    % (shown with dashed black arrows only for OFF).
    % The region $\mathcal{R}$1 around NCG~1275 (defined by $\theta<\theta_{\mathrm{min}}$ with respect to NGC~1275's direction,
    % which is coincident with the center of the cluster), is used to evaluate its
    % gamma-ray emission activity for each given dataset.}
\end{figure}
MAGIC observations were carried out in wobble mode~\citep{Fomin:1994aj}, where both the signal (ON) and background control (OFF) regions are 
observed within the same field of view (FoV). 
The data were taken in two different observation pointing modes (here labelled $\mathcal{A}$ and $\mathcal{B}$).
The gamma-ray emitting radio-galaxy NGC~1275 is located at the dynamical center 
%\textbf{(well defined?)} 
of the cluster (see~\autoref{fig:perseusFOV}), 
and for observation mode $\mathcal{A}$, four symmetric pointing positions are taken 
at $0.4^{\circ}$ distance around this point. 
In pointing mode $\mathcal{B}$, 
%the center of the Perseus CG is kept in the FoV
%while also monitoring IC~310 (another gamma-ray emitter member of the cluster).
%In this $\mathcal{B}$ mode, 
the instrument wobbles around a point half-distance between NGC~1275 and IC~310
(pointing alternately in two of the pointing positions of mode $\mathcal{A}$).
%For each observation mode ($\mathcal{A}$ or $\mathcal{B}$), ON/OFF regions from opposing pointings are analyzed together.
The galaxy NGC~1265 is another important object in the FoV.
NGC~1265 is clearly visible in X-rays~\citep{Sun:2005yd}
and, albeit never detected above $E>1~\GeV$,  
is treated as a potential gamma-ray emitter in the analysis.
\bigskip

During the observation campaign, the MAGIC telescopes underwent several hardware
upgrades~\citep{Aleksic:2014lkm,Aleksic:2014poa}, 
leading to six different \emph{hardware stable periods} (from $\mathcal{P}1$ to $\mathcal{P}6$ in~\autoref{tab:data}).
Appropriate Monte Carlo (MC) simulations for each period are generated to determine the corresponding instrument response function (IRF 
i.e., the effective area for signal, the angular resolution and bias of the energy reconstruction). 
\bigskip

For each data sample, the standard MAGIC event reconstruction~\citep{Aleksic:2011bx} is applied.
Data selection is performed in two different steps, first based on \emph{quality} cuts and secondly on \emph{specific} cuts (see~\autoref{tab:data} for details on the amount of data surviving each data selection criteria).
Quality cuts are used to select data runs of $\sim$20 minutes duration with the zenith angle ranging between $5^{\circ}$ and $50^{\circ}$. 
Only a minor fraction of the data recorded was taken with zenith angles above $50^{\circ}$. 
A second quality cut was based on the intensity of the night sky background (NSB) that, if too large, also significantly reduces the performance. 
We allowed the average camera illumination to be no larger than three times that of a standard dark night~\citep[as suggested in][]{Ahnen:2017vsf}.
Furthermore, we selected data based on atmospheric transparency measured with the MAGIC LIDAR
instrument~\citep{Fruck:2014mja}, 
requiring a atmospheric optical depth in the direction of the telescope pointing larger than 85\% that of a clear 
night~\citep[which guarantees acceptable performance and 
systematics below those quoted in][]{Aleksic:2014lkm}. 
Finally, an event-wise cut based on the size of the event (the total integrated charge contained in a shower image) of $80$ 
photo-electrons is applied. 
This is slightly higher than the one used for standard low zenith observations to compensate for the larger extinction of 
Cherenkov light from events at higher zenith values present in our data.
%Data selection is performed in two different steps, \emph{quality} and \emph{specific} cuts (see~\autoref{tab:data} for details on the amount of data surviving each data selection criteria).
%Quality cuts are used to select data runs ($\sim20$ min long each) with the zenith angle ranging between $5^{\circ}$ and $50^{\circ}$. 
%We further selected data runs based on the NSB light level, allowing the average camera
%illumination to be
%no larger than three times that of a standard dark night~\citep[as suggested in][]{Ahnen:2017vsf}.
%to guarantee maximum performance.
%During all observations, LIDAR~\citep{Fruck:2014mja} measurements were taken in parallel and data 
%with aerosol transparency between the telescopes and 9 km above ground larger than 85\% at the time of observation~\citep[which guarantees acceptable performance and systematics below those quoted in][]{Aleksic:2014lkm} are selected for further analysis. 
%We apply strong cuts in NSB and LIDAR values, stronger than the ones used in other analyses within the same dataset, 
%which explain the low efficiency of the selected data.
%Finally, an event-wise cut based on the size of the event 
%(the total integrated charge contained in a shower image) of $80$ photo-electrons is applied.
%The size cut is selected to be slightly higher than the one used for standard low/middle zenith observations (zd~$<35^\circ$)
%to compensate for the larger extinction of Cherenkov light from events at higher zenith values. 
\bigskip

In a second step, specific cuts are used to remove observation nights in which the detection significance~\citep[defined in ][]{LiMa:1983ApJ} of any of the astrophysical sources NCG~1275, NGC~1265, and IC~310 
(colored star markers in~\autoref{fig:perseusFOV}) 
is higher than 3$\sigma$. 
The gamma-ray emission of these sources may vary from night to night both in intensity and spectral morphology.
Since we search for a steady signal of DM, 
excluding from the search data from these nights 
minimizes possible systematic effects introduced by astrophysical signal contamination.
No bias is introduced in the search for DM since the evaluation of the detection significance 
of NGC~1275, NGC~1265 and IC~310 is performed out of the signal region used for the decaying DM search.
\newline

Finally, events surviving all the aforementioned data selection criteria are assigned 
an estimated energy and direction, and a parameter called "hadronness" or $h$~\citep[based 
on a random forest method, as explain in][]{Albert:2007yd},
which estimates the hadron or gamma-ray origin of an event. 
\newline

%\paragraph{Dark Matter signal region}
\subsection{Dark Matter signal region}
%According to the DM density profile described in~\autoref{sec:dm}, the virial radius of the Perseus CG is seen under an angle of $\sim1.5^\circ$. %~\citep{Matsushita:2013ur}
%which is 
%(much larger than the angular resolution of MAGIC). 
For an accurate computation of the IRF of the analysis, 
the morphology of the expected DM signal (described in~\autoref{sec:dm}) is used to 
tune the distribution of simulated MC events. % used to compute the IRFs of the analysis. 
This procedure was first applied by~\cite{Ahnen:2017pqx}
during the DM search from the Ursa Major II dSph
and was discussed extensively in~\citet{Palacio:thesis}.
Moreover, in order to %search for an extended signal of decaying DM while 
avoid contamination from gamma rays coming from NGC~1275, 
%(the region where the astrophysical emission is expected to dominate over the DM signal), 
we construct a \emph{ring-shaped} signal-search region $\mathcal{R}2$
(defined by two angular distances $\theta_{\mathrm{min}}$ and $\theta_{\mathrm{max}}$ described in~\autoref{fig:perseusFOV}).
Apart from excluding the location of NGC~1275 from the region of interest $\mathcal{R}2$, its astrophysical contamination  
inside $\mathcal{R}2$ due to miss-reconstructed events is also estimated and included in the analysis.
Due to the large extension of the decay DM signal and the finite distance between ON and OFF regions
%(due to the pointing strategy),
(regions around black and empty stars in~\autoref{fig:perseusFOV}),
%the shape of the ON and OFF regions, 
OFF regions are not fully signal-free
(in other words, DM events are expected inside the OFF region for all pointing directions).
This contamination is also taken into account
in the likelihood ($\lkl$, see~\autoref{sec:BinnedLikelihoodAnalysis}) and is estimated to be of $\sim10$\% ($\sim40$\%) 
of the signal integrated in ON for pointing mode $\mathcal{A}$ ($\mathcal{B}$), a factor that further affects the sensitivity.

%\textbf{(move to later?)\md{no}}
%\paragraph{Gamma-ray shower discrimination and final event selection}
\subsection{Gamma-ray shower discrimination and final event selection}
\label{subsec:eventSelecion}
As seen in \autoref{eq:diffflux} and \ref{eq:PPandJfactor}, the expected gamma-ray flux depends on the instrument-related parameters $\theta_{\mathrm{min}}$ and $\theta_{\mathrm{max}}$ and
%(based on the beginning of this section) 
also depends on the cut on hadronness $h_{\mathrm{c}}$,
optimized independently in each energy bin.
%applied in energy dependent bins based on the fraction of MC events that survive each cut).
Both cuts are optimized based on their \emph{expected sensitivity to the 
DM decay lifetime}
%($\tdm^{\mathrm{svt}}$), approximated as $\tdm^{\mathrm{svt}}=\left(1/\tdm^{\text{LL}}-\widehat{1/\tdm}\right)^{-1}$ 
(as will be introduced in~\autoref{sec:BinnedLikelihoodAnalysis}).
The optimal selection values are then $h_{\mathrm{c}}$, selected for a MC event efficiency of 80\%, $\theta_{\mathrm{min}} = 0.1^\circ$ and 
$\theta_{\mathrm{max}} = 0.33^\circ$ (note that $\theta_{\mathrm{max}}$ is already close to 0.4$^{\circ}$, 
the wobble distance at which data were taken).
%\newline
The effective $J$-factor for those cuts\footnote{Estimated by $J_\text{dec}\cdot{N^{\theta_{\text{min}}<\theta<\theta_{\text{max}}}}/{N^{\theta<1.5^{\circ}}}$, 
with $N^{\theta<1.5^{\circ}}$ and $N^{\theta_{\text{min}}<\theta<\theta_{\text{max}}}$ the number of simulated gamma-ray events, 
following the spatial distribution expected for the decay DM signal from Perseus CG, detected before and after the angular cuts, 
respectively.} is $\sim0.99\times10^{18}~\mbox{GeV}\,\mbox{cm}^{-2}$.
\section{Dark matter decay search}
\label{sec:BinnedLikelihoodAnalysis}
\begin{figure}[t!]
  \centering
\includegraphics[width=0.7\linewidth]{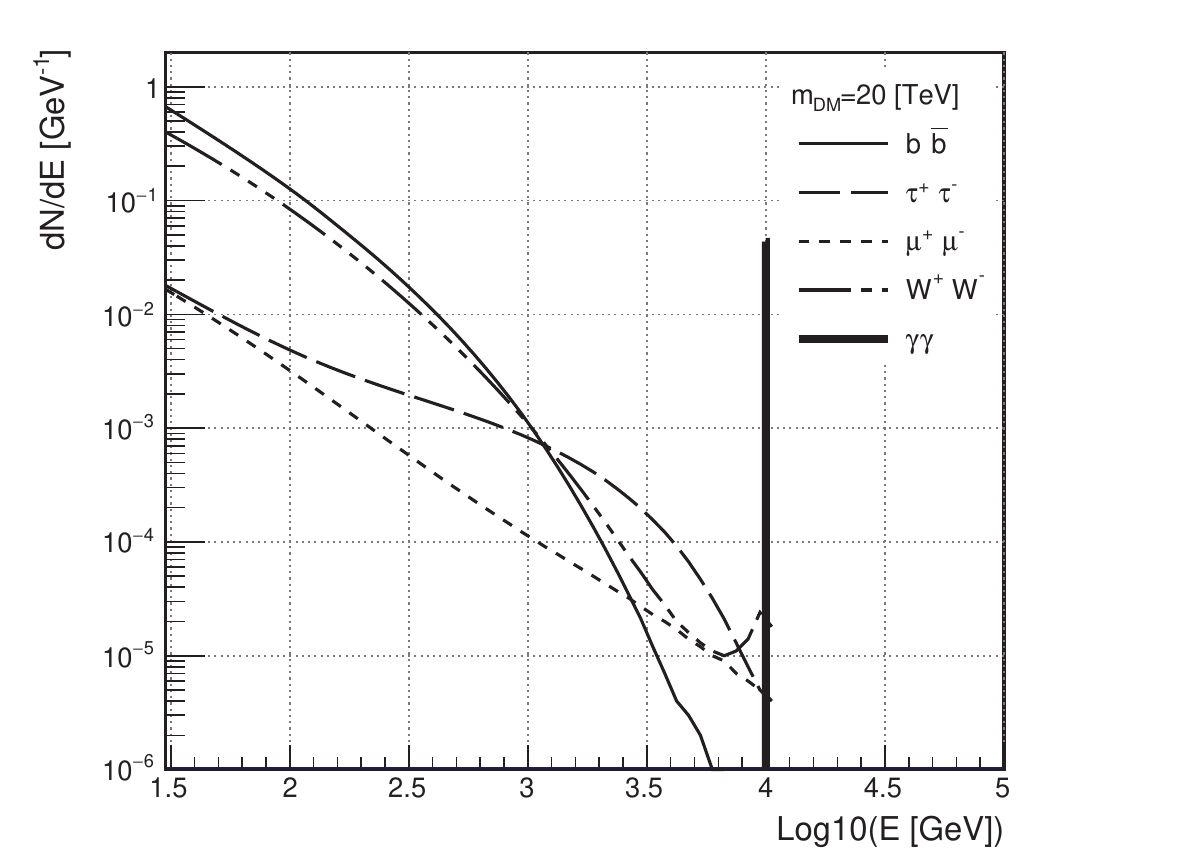}
  \caption{\label{fig:dmdecay} Average gamma-ray spectra (${dN}/{dE}$) 
  as a function of the gamma-ray energy ($E$) 
  due to prompt emission for the decay
  channels $b\bar{b}$, $\mu^+\mu^-$, $\tau^+\tau^-$, $W^+W^-$ and $\gamma\gamma$ 
  for a 20~TeV DM particle.}
\end{figure}
\noindent
%, a dedicated analysis for indirect DM searches with IACTs that
%takes advantage of the distinct features expected in the gamma-ray spectrum of DM origin.
%The method, used in~\cite{Aleksic:2013xea}, lead to the most stringent constraints
%to the annihilation cross-section of DM particles for masses above few TeV.
Following \cite{Ahnen:2016qkx}, we use the \texttt{PYTHIA} simulation package 
version \texttt{8.205}~\citep{Sjostrand:2014zea} to 
%model the gamma-ray DM gamma-ray emission.
%We 
compute the average gamma-ray spectrum per decay process (${\mathrm{d}N}/{\mathrm{d}E}$) for DM particles of masses between
$200$~GeV and $200$~TeV decaying into the SM pairs $b\bar{b}$, $\tau^+\tau^-$, $\mu^+\mu^-$, $W^+W^-$ and $\gamma\gamma$.
For each channel and mass, we average the gamma-ray spectrum resulting from $10^{7}$ decay events of a generic resonance with mass $m_{DM}$ 
into the considered pair (see~\autoref{fig:dmdecay}). 
For each simulated event, we trace all the decay chains, including the muon radiative decay 
($\mu^{-}\rightarrow e^{-}\overline{\nu}_{e}\nu_{\mu}\gamma$, not active in PYTHIA by default), 
down to stable particles. % (see~\autoref{fig:dmdecay}).
To search for DM in the Perseus CG, we use a binned likelihood method developed for indirect DM searches with
IACTs~\citep{Aleksic:2012cp}.
\newline

The binned likelihood used in our analysis is written as
\begin{eqnarray}\label{eq:binnedlikelihood}
\nonumber 
 &&\lkl \left(1/\tdm;\boldsymbol{\bm{\nu}}\, |\, \boldsymbol{\data}\right) \\
%
%
%\nonumber
%&=& \lkl \left(1/\tdm; \{\bij\}_{i=1,\dots,N_{\text{samples}};\,j=1,\dots,\Nbins}\, , \tau_i\, , \mathcal{J}|\, 
% ( \Nonij,\Noffij)_{i=1,\dots,N_{\text{samples}};\,j=1,\dots,\Nbins} \right)  \\
%
%
%\nonumber
%&=& \mathcal{J}(J|\Jobs) \\
\nonumber 
&=& \prod_{i=1}^{N_{\text{samples}}}
\mathcal{K}(\kappa_i|\tauobsi,\sigmataui) \\
%& & \mathcal{T}(\tau_i|\tauobsi,\sigmataui) \\
\nonumber
& & \times \prod_{j=1}^{\Nbins} 
\biggl[ \frac{\left(\gij(\tdm) +\bij + \fij\right)^\Nonij}{\Nonij!} e^{-\left(\gij(\tdm)+\bij+\fij\right)} \\
& & \times\,
\frac{\left(\kappa_i \bij + \goffij(\tdm)\right)^\Noffij}{\Noffij!} e^{-\left(\kappa_i\bij+\gij^{\text{OFF}}(\tdm)\right)} \biggr], 
%\nonumber \\
\end{eqnarray}
where $\bm{\nu}$ collectively refers to the nuisance parameters and $\data$ to the data being
\begin{eqnarray}
 %&& \boldsymbol{\bm{\nu}} = \{\bij\}_{i=1,\dots,N_{\text{samples}};\,j=1,\dots,\Nbins}\, , \kappa_i \\
  && \boldsymbol{\bm{\nu}} = \{\bij\} \, , \kappa_i \\
\nonumber
% && \boldsymbol{\data} = ( \Nonij,\Noffij)_{i=1,\dots,N_{\text{samples}};\,j=1,\dots,\Nbins}.
  && \boldsymbol{\data} = ( \Nonij,\Noffij).
\end{eqnarray}
\noindent
The index $i$ refers to the independent datasets
(described in~\autoref{tab:Lkl_bins}), and $j$ to the bins of estimated energy. 
%The nuisance parameters are collectively referred as $\bm{\nu}$ and $\data$ represents the datasets. 
The parameters $\gij$ and $\goffij$ are the estimated number of DM signal events for the ON and OFF regions, respectively; 
the parameters $\bij$ are the estimated number of background events;
% (see~\ref{sec:Scaling}),
$\Nonij$ are the number of observed events in the ON region and $\Noffij$ is the number of observed events in the
corresponding OFF bin; 
%$\mathcal{J}$ is the likelihood for the $J$-factor
%(currently being $\mathcal{J}=\delta\left(J-\Jobs\right)$)\footnote{
%Following~\cite{Ahnen:2016qkx}, we also show the result for the case of $\mathcal{J}$ being 
%a Gaussian function in $J$, centered in $J_{obs}$, with $60\%$ uncertainty ($\sigma=0.6\,J_{obs}$).}; 
$\mathcal{K}$ is the likelihood function for
$\kappa_i$ (the OFF/ON acceptance ratio), parametrized by a Gaussian
function with mean $\tauobsi$ and variance $\sigmataui^2$,
which includes statistical and systematic uncertainties, added in quadrature assuming Poisson statistics. 
We consider a systematic uncertainty for the parameter $\kappa_i$, $\sigmatausys=0.015\tauobsi$, 
a value that has been established in~\citep{Aleksic:2014lkm}.
%At high statistic ($>10^4$ ON events $\sim50$ h), the statistical error of $\tau$ goes beyond
%the systematic wall (how good do we perform at estimating the background rate at a given region)
%\newline
$\fij$ (considered as fixed parameters in this analysis, in order not to lose the convergence of the likelihood) 
are the estimated number of foreground events from NGC~1275 
(computed by extrapolating from the estimated number of gamma-ray excess-events within $\mathcal{R}1$ around NGC~1275,
see~\autoref{fig:perseusFOV}). 
%using the spatial distribution of a point-like MC.
%In order to model NGC~1275 contamination into $\mathcal{R}2$
%we measure the number of NGC~1275 excess
%events inside $\mathcal{R}1$ ($f_{\mathcal{R}1}$), a circular region of radius $\theta_{min}$ centered on NGC~1275 
%(which is also the center of the Perseus CG).
We infer the number of expected events reconstructed
inside $\mathcal{R}2$ taking into account the instrument's angular point spread function
computed from a point-like MC representative for each analyzed data sample~\cite[same hardware stable period, 
same zenith range, and weighted to reproduce NGC1275's spectra measured in][]{Ahnen:2016qkt}.
$\bij$ and $\kappa_i$ are nuisance parameters,
whereas the estimated number of signal events $\gij$ and $\gij^{OFF}$ depend on the free parameter $\tdm$ through 
\begin{eqnarray}
\label{eq:freeparameter}
 \gij(\tdm) &=& \Tobsi \int_{\epMinj}^{\epMaxj} \mathrm{d}E'
\int_0^\infty \mathrm{d}E
\frac{\mathrm{d}\phi(\tdm)}{\mathrm{d}E}\, \aeff(E)_{i}\, G(E'|E)_{i}, \\
%\nonumber
%\gij^{\text{OFF}}(1/\tdm) &=& \gij^{\text{OFF}}\left(\gij(1/\tdm)\right)  \\
\label{eq:freeparameterOff}
\gij^{\text{OFF}}(\tdm) &=&\Tobsi \int_{\epMinj}^{\epMaxj} \mathrm{d}E'
\int_0^\infty \mathrm{d}E
\frac{\mathrm{d}\phi(\tdm)}{\mathrm{d}E}\, \aeff(E)_{i}\, \epsilon(E)_i \, G(E'|E)_{i}. 
\end{eqnarray}
$\Tobsi$ is the total observation time, $E$ and $E'$ the true and estimated gamma-ray energy respectively, 
%respectively, 
and $\epMinj$ and $\epMaxj$ the minimum and maximum energies 
%respectively, 
of the $j$-th energy bin.  
Finally, $G$ is the probability density function for the energy estimator $E'$ for true energy $E$, 
$\aeff$ is the effective collection area for $\mathcal{R}2$ angular cuts computed from the tailored
MC sample introduced in~\autoref{sec:data} 
(that takes into account the expected morphology of the gamma-ray signal and the instrument angular
resolution), 
and $\epsilon_i$, the ratio between expected number of signal events in the OFF and ON regions, obtained from the same MC sample.
%$\aeff$ 
%and $\aeff^{\text{OFF}}$ 
%is the effective collection area for $\mathcal{R}2$.
%$\epsilon$ (in~\autoref{eq:freeparameterOff}) parametrizes the efficiency of gamma rays coming from decaying DM
%in Perseus to be reconstructed within OFF and is computed from the taylored MC introduced in~\autoref{sec:data}
%(that takes into account the expected morphology of the gamma-ray signal and the instrument angular resolution).
\newline

%In order to optimize the signal search strategy with respect to these quantities, 
We define the profile likelihood ratio as
\begin{equation}\label{eq:001_IACT_profLklRatio}
\lambda_{P}\left(1/\tdm|\, \data\right) = 
%\frac{\lkl (1/\tdm;\widehat{\widehat{\bm{\nu}}}\, |\, \data) }{\lkl (\widehat{{1}/{\tdm}} ;\widehat{\bm{\nu}}\, |\, \data) },
\frac{\lkl (1/\tdm;\doublehatnu\, |\, \data) }{\lkl (\widehat{{1}/{\tdm}} ;\hatnu\, |\, \data) },
\end{equation}
where
$\widehat{1/\tdm}$ and $\hatnu$ are the values maximizing $\lkl$
($\lkl$ is linear in $1/\tdm$), and $\doublehatnu$ the 
value that maximizes $\lkl$ for a fixed $1/\tdm$ (when performing the maximization %\cite{Fischler:2003rk},
we restricted the value of the lifetime to the physical range, $1/\tdm\geq$0).
Lower limits in $\tdm$ at 95\% CL ($\tdm^{\text{LL}}$) are given for 
\begin{equation}
-2\ln\lambda_{P}\left(1/\tdm^{\text{LL}} \; \vline \; \data \right) = 2.71. 
\end{equation}
We approximate the expected sensitivity to the DM decay lifetime $\tdm^{\mathrm{svt}}$ as
\begin{equation}
\tdm^{\mathrm{svt}}=\left(1/\tdm^{\text{LL}}-\widehat{1/\tdm}\right)^{-1}.
\end{equation} 
%where we compute $-2\ln\lp$ vs.\ $1/\tdm$ for the range of $1/\tdm$ fulfilling $-2 \ln \lp(1/\tdm) \leq 2.71$. 
%
% \autoref{fig:NGC1275dNdEp} shows the measured $\gammaaray$ event rate from NGC~1275 ($dN/dE'$) in the region $\mathcal{R}1$.
%
% \begin{figure}
%   \centering
%   \includegraphics[width=0.7\linewidth]{figure/expecteddNdEpNGC1275_250h_allMonteCarloPeriods.pdf}
% \caption{NGC~1275's measured event rate ($dN/dE'$, being $E'$ the reconstructed energy) in the region $\mathcal{R}1$,
% for both observational projects ($\mathcal{A}$ and $\mathcal{B}$),
% for all HSP (ST.01.02, ST.03.01, ST.03.03, ST.03.05, ST.03.06, ST.03.07) 
% and both zenith ranges (\emph{05to35} to \emph{35to50}).
% Each data sample has been normalized to $250$ h, so that comparisons between them can be performed.}
% \label{fig:NGC1275dNdEp}
% \end{figure}
%\newline

The null hypothesis is the case with no DM signal ($1/\tdm=0$), 
while the test hypotheses are built considering the flux
computed using \autoref{eq:PPandJfactor}, under the assumption of different DM particles with masses from $200~\GeV$ to
$200~\TeV$ for pure SM decays. 
The dataset is divided into $N_{\text{samples}}=42$ independent subsets\footnote{ 
Four subsamples, out of the naively expected 44 from \autoref{tab:Lkl_bins}
%(6 periods x 1 zenith band x 4 pointings = 24 for pointing $\mathcal{A}$, plus 5 x 2 x 2 = 20 for $\mathcal{A}$) 
were excluded from the search since almost no data survived the cuts introduced in \autoref{tab:data}.
Moreover, two extra pointings (included in \autoref{tab:data} but not described in \autoref{tab:Lkl_bins}) 
were taken at the beginning of the campaign at the same wobble distance but with different orientation, 
before optimizing pointing mode $\mathcal{A}$ and $\mathcal{B}$.
These two subsamples (accounting each of them for $\sim$17~h) are also included in the analysis. 
%Data was  being W5=+000 and W6=+180.
} according to the two 
observational pointing schemes ($\mathcal{A}$ and $\mathcal{B}$), the wobble pointing positions,
the different hardware stable periods, and two zenith ranges (15-35$^\circ$ and 35-50$^\circ$).
%leading to 42 terms in the likelihood ($N_{\text{samples}}=42$ in~\autoref{tab:Lkl_bins}, where 
Each likelihood term is linked to the rest of the terms through the common physical parameter $\tdm$.
\begin{table}
    \centering
    % \begin{tabular}{c|c}
    %     % \multicolumn{1}{l}{NGC-1275} \\$\mathcal{A}$
    %     \multicolumn{1}{l}{$\mathcal{A}$} \\
    %     \hline
    %     \hline
    %      HSP& ST.02.01, ST.03.01, ST.03.03,\\
    %      &  ST.03.05, ST.03.06, ST.03.07\\
    %      \hline
    %      Zenith angle&  $5-35$\\
    %      \hline
    %      Wobble pointing & $W0.4\left(+058,+157,+238,+337\right)$ \\
    %     \hline
    %     \hline
    %     \multicolumn{1}{l}{}\\
    %     % \multicolumn{1}{l}{PERSEUS-MA}\\
    %     \multicolumn{1}{l}{$\mathcal{B}$} \\
    %     \hline
    %     \hline
    %      HSP& ST.03.01, ST.03.03, ST.03.05\\
    %      &  ST.03.06, ST.03.07\\
    %      \hline
    %      Zenith angle&  $5-35$, $35-50$\\
    %      \hline
    %      Wobble pointing & $W0.26\left(+108,+288\right)$  \\
    %     \hline
    %     \hline
    % \end{tabular}
    \begin{tabular}{c|c}
        % \multicolumn{1}{l}{NGC-1275} \\$\mathcal{A}$
        \multicolumn{1}{l}{$\mathcal{A}$} \\
        \hline
        \hline
         Hardware stable period & $\mathcal{P}1$, $\mathcal{P}2$, $\mathcal{P}3$, $\mathcal{P}4$, $\mathcal{P}5$, $\mathcal{P}6$\\
         \hline
         Zenith angle&  [$5$-$35$]\\
         \hline
         Wobble pointing & $W0.4\left(+058^\circ,+157^\circ,+238^\circ,+337^\circ\right)$ \\
        \hline
        \hline
        \multicolumn{1}{l}{}\\
        % \multicolumn{1}{l}{PERSEUS-MA}\\
        \multicolumn{1}{l}{$\mathcal{B}$} \\
        \hline
        \hline
         Hardware stable period & $\mathcal{P}2$, $\mathcal{P}3$, $\mathcal{P}4$, $\mathcal{P}5$, $\mathcal{P}6$\\
         \hline
         Zenith angle&  [$5$-$35$], [$35$-$50$]\\
         \hline
         Wobble pointing & $W0.26\left(+108^\circ,+288^\circ\right)$  \\
        \hline
        \hline
    \end{tabular}
    \caption{List of the different bins for which independent IRFs (and hence, independent likelihoods) are defined.}
    \label{tab:Lkl_bins}
\end{table}
% All different subsamples are analysed using specific IRFs.
%\newline
%
%
%\noindent
%The profile likelihood ratio is defined as:
%\begin{equation}\label{eq:profLklRatio}
%\lambda_{P}\left(1/\tdm|\, \boldsymbol{\data}\right) = 
%\frac{\lkl (1/\tdm;\boldsymbol{\widehat{\widehat{\bm{\nu}}}}\, |\, \boldsymbol{\data}) }{\lkl (1/\widehat{\tdm};\widehat{\bm{\nu}}\, %|\, \boldsymbol{\data}) },
%\end{equation}
%where $\widehat{\tdm}$ and $\boldsymbol{\widehat{\bm{\nu}}}$ are the values that maximize $\lkl$, 
%and $\widehat{\widehat{\bm{\nu}}}$ the value that 
%%maximises $\lkl$ for a given $\tdm$.
%Lower limits in $\tdm$ at 95\% CL ($\tdm^{LL}$)  are given for: 
%\begin{equation}
%-2\ln\lambda_{P}\left(1/\tdm^{LL} \; \vline \; \boldsymbol{\data} \right) = 2.71. 
%\end{equation}
%When running the minimisation %\cite{Fischler:2003rk},
%we restricted the value of the lifetime to the physical range ($1/\tdm\geq$0).

%\newpage
\section{Results and Discussion}
\label{sec:results}
\begin{figure}[ht!]
  \centering
  \includegraphics[width=.45\linewidth]{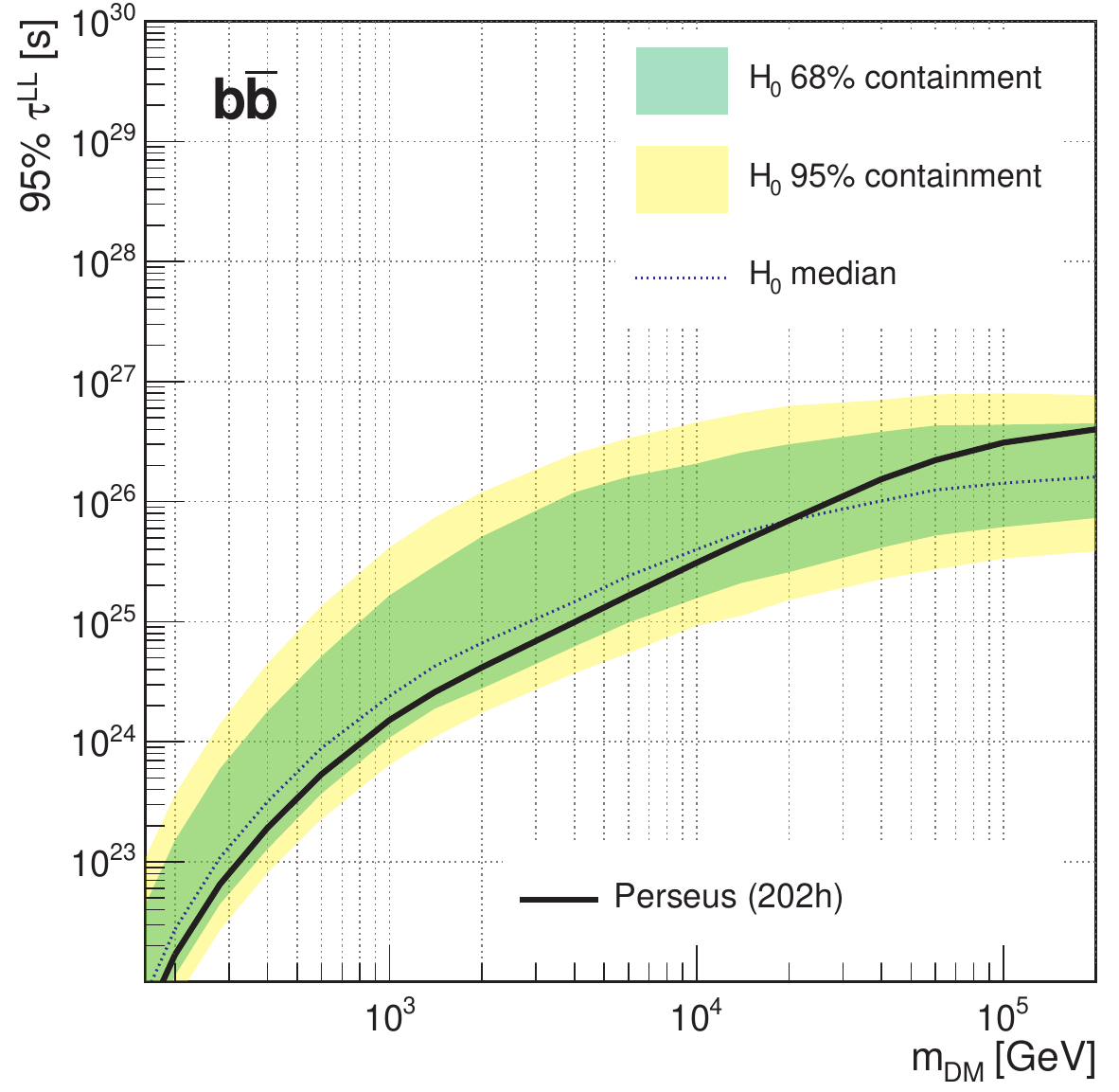}
  \includegraphics[width=.45\linewidth]{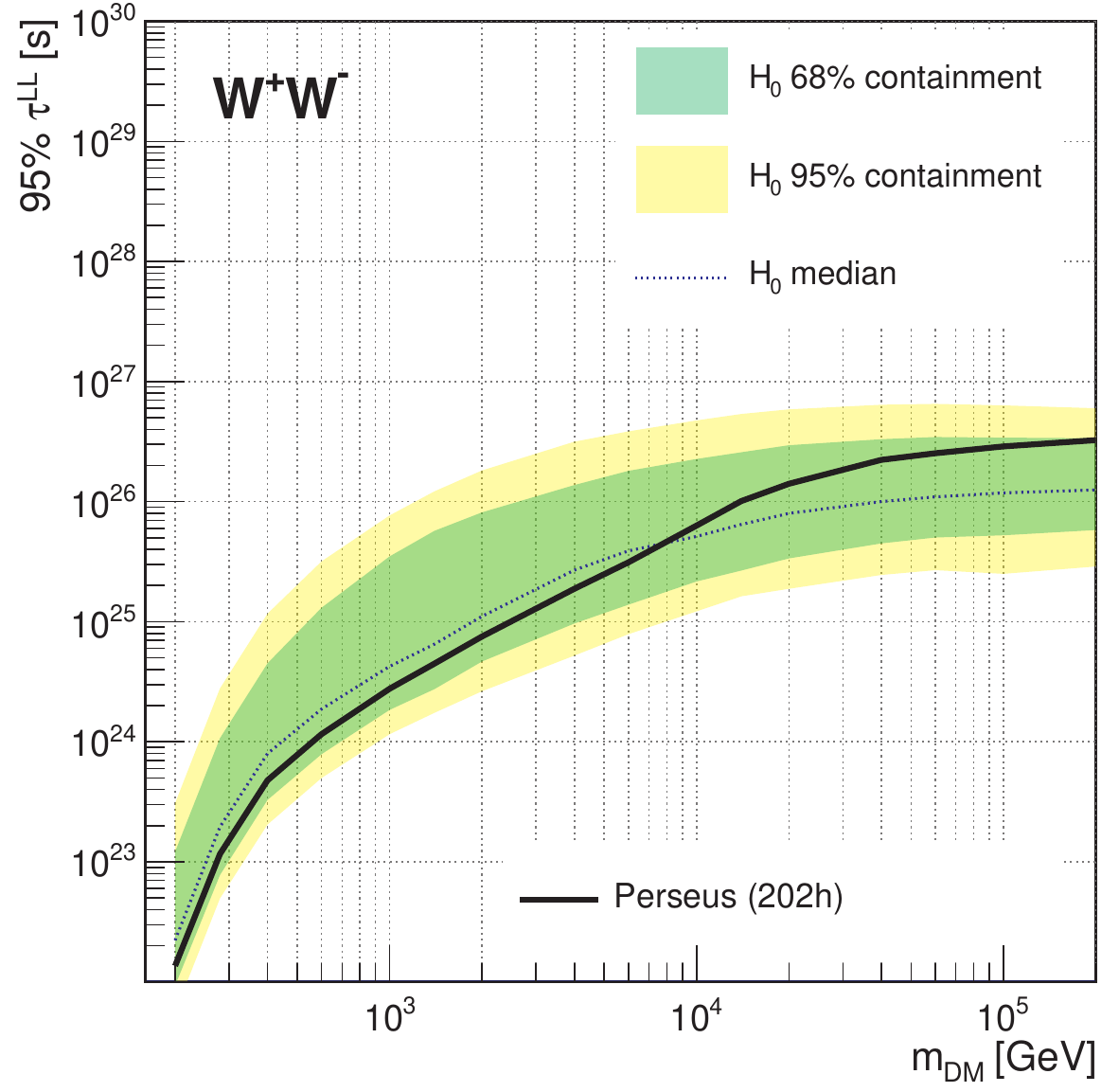}
  \includegraphics[width=.45\linewidth]{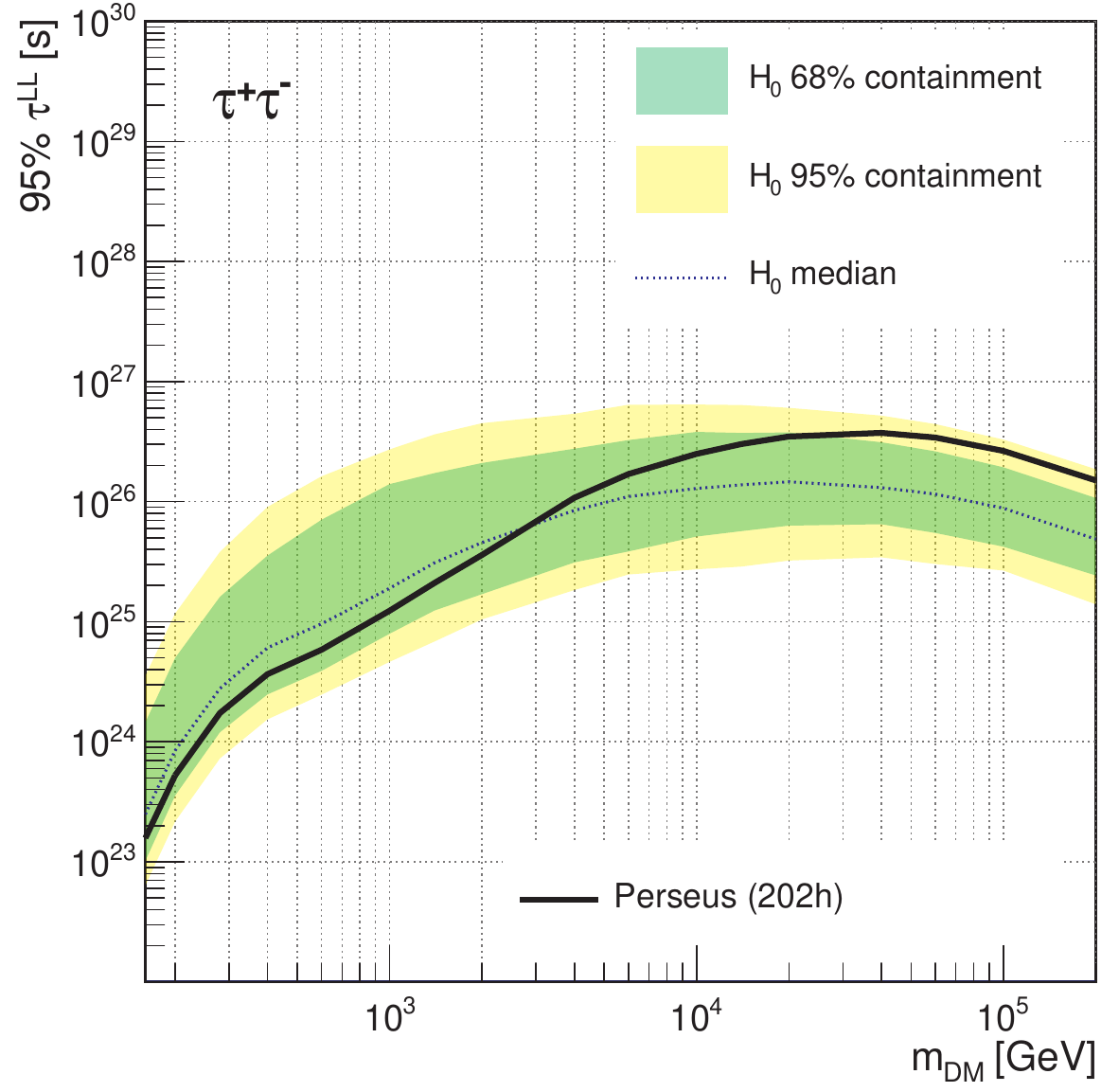}
  \includegraphics[width=.45\linewidth]{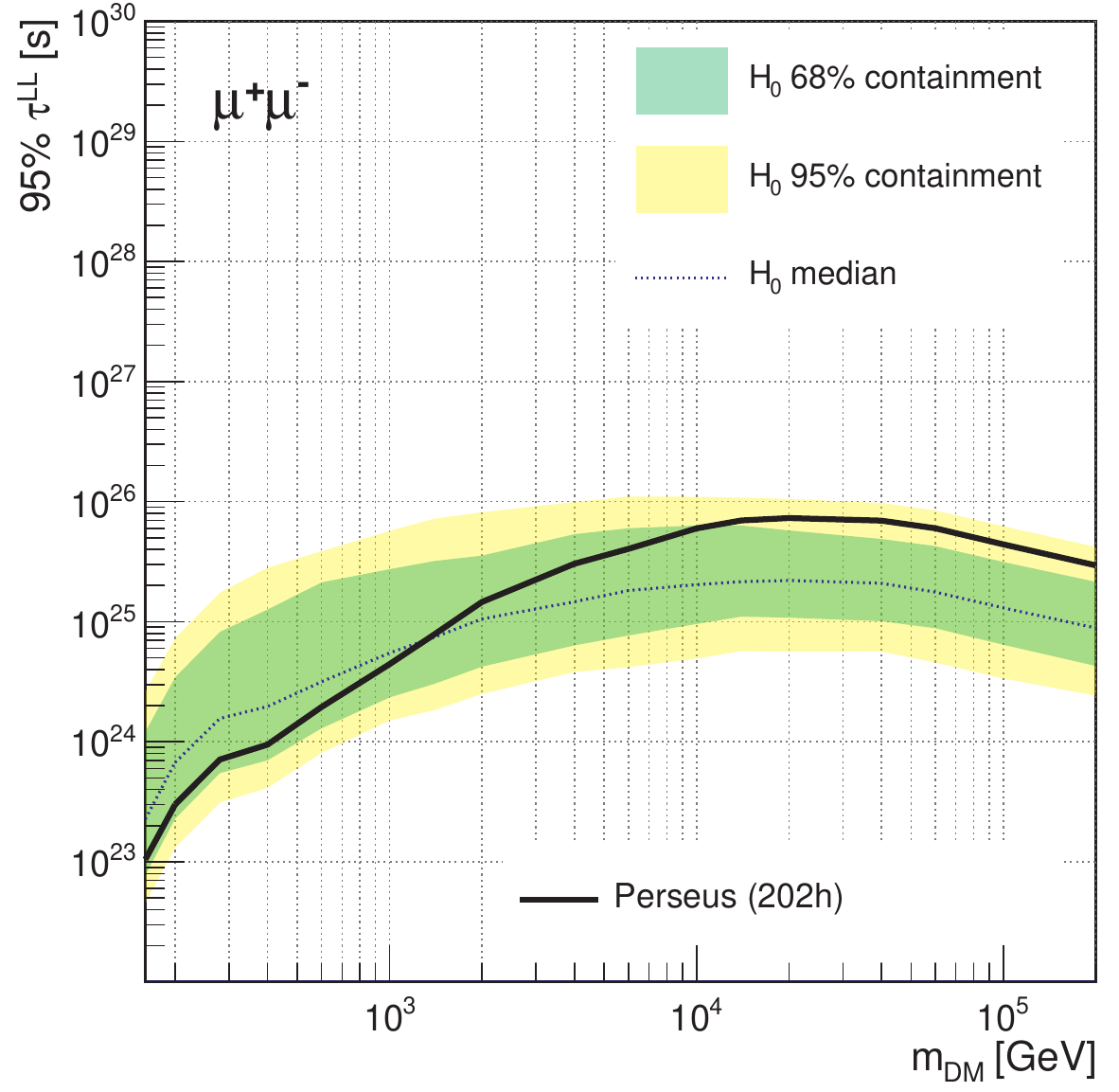}
\caption{\label{fig:results_continuous}
95\% CL lower limit on the DM decay lifetime (solid line) in the $b\bar{b}$~(top-left), $W^{+}W^{-}$~(top-right), $\tau^{+}\tau^{-}$~(bottom-left) and $\mu^{+}\mu^{-}$~(bottom-right) channels using 202~h of Perseus CG data.
The expected limit (dashed line) and the two sided $68\%$ and $95\%$ containment bands are also shown.}
\end{figure}
\begin{figure}[ht!]
\centering
  \includegraphics[width=.45\linewidth]{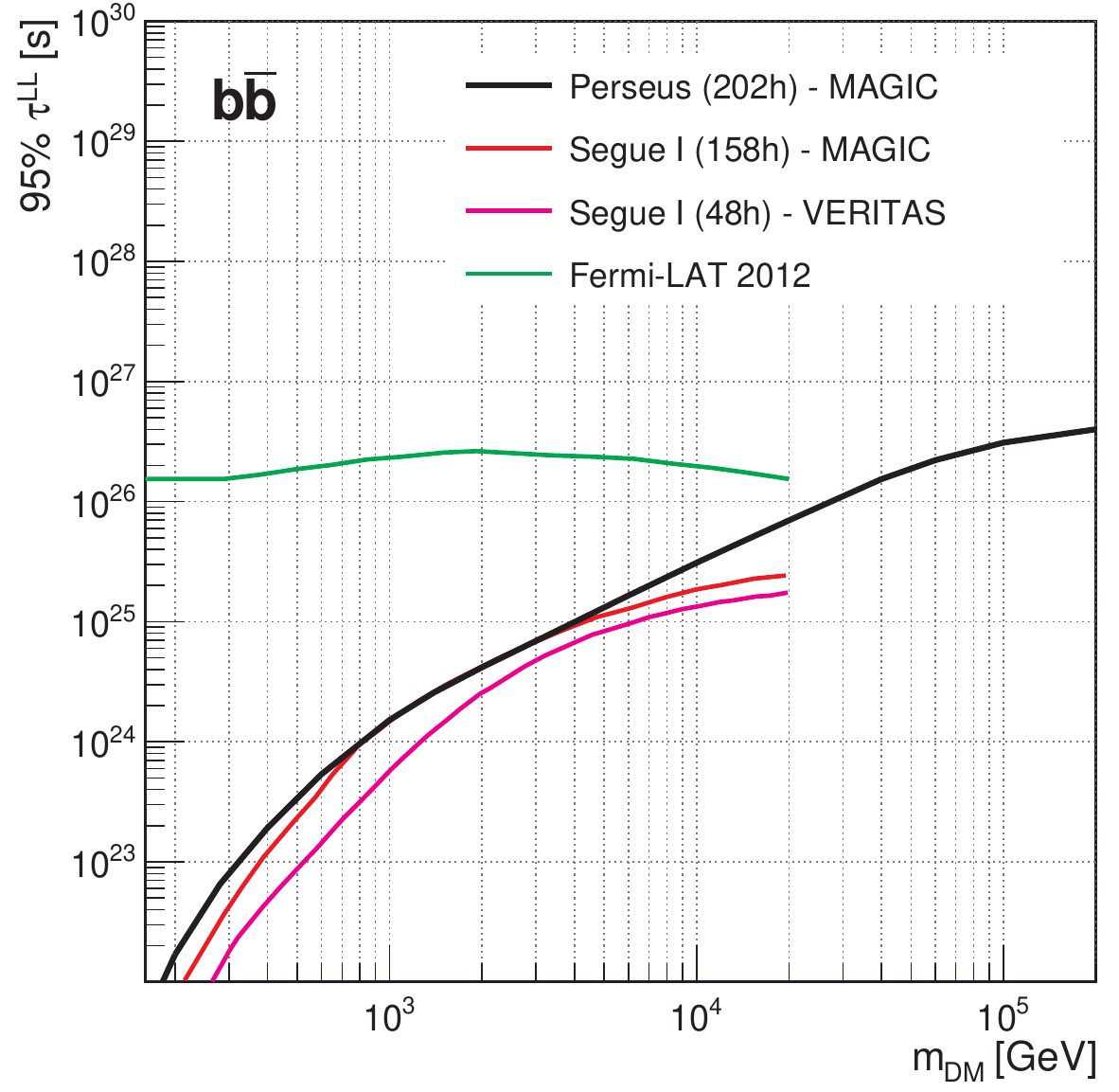}
  \includegraphics[width=.45\linewidth]{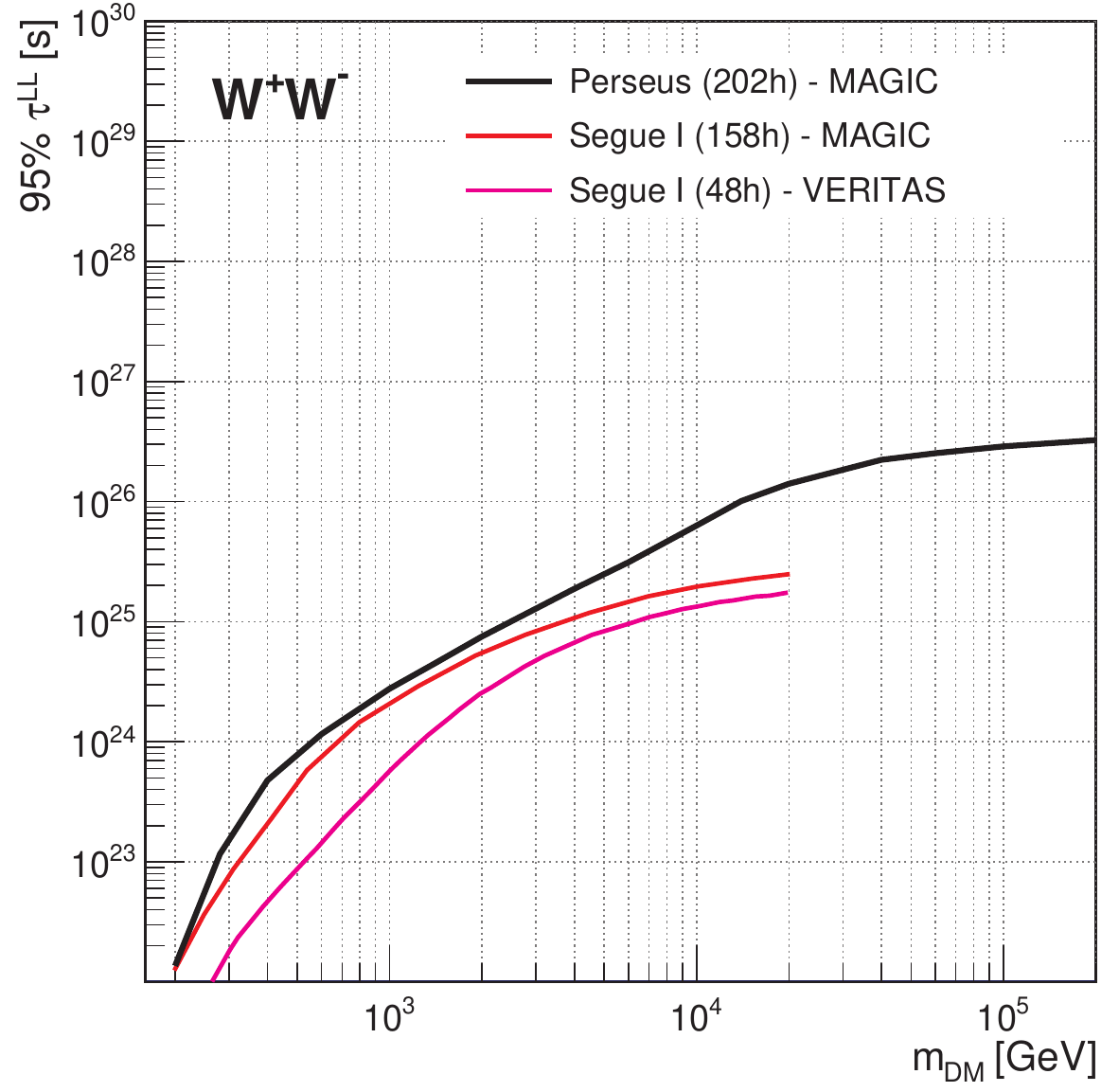}
   \includegraphics[width=.45\linewidth]{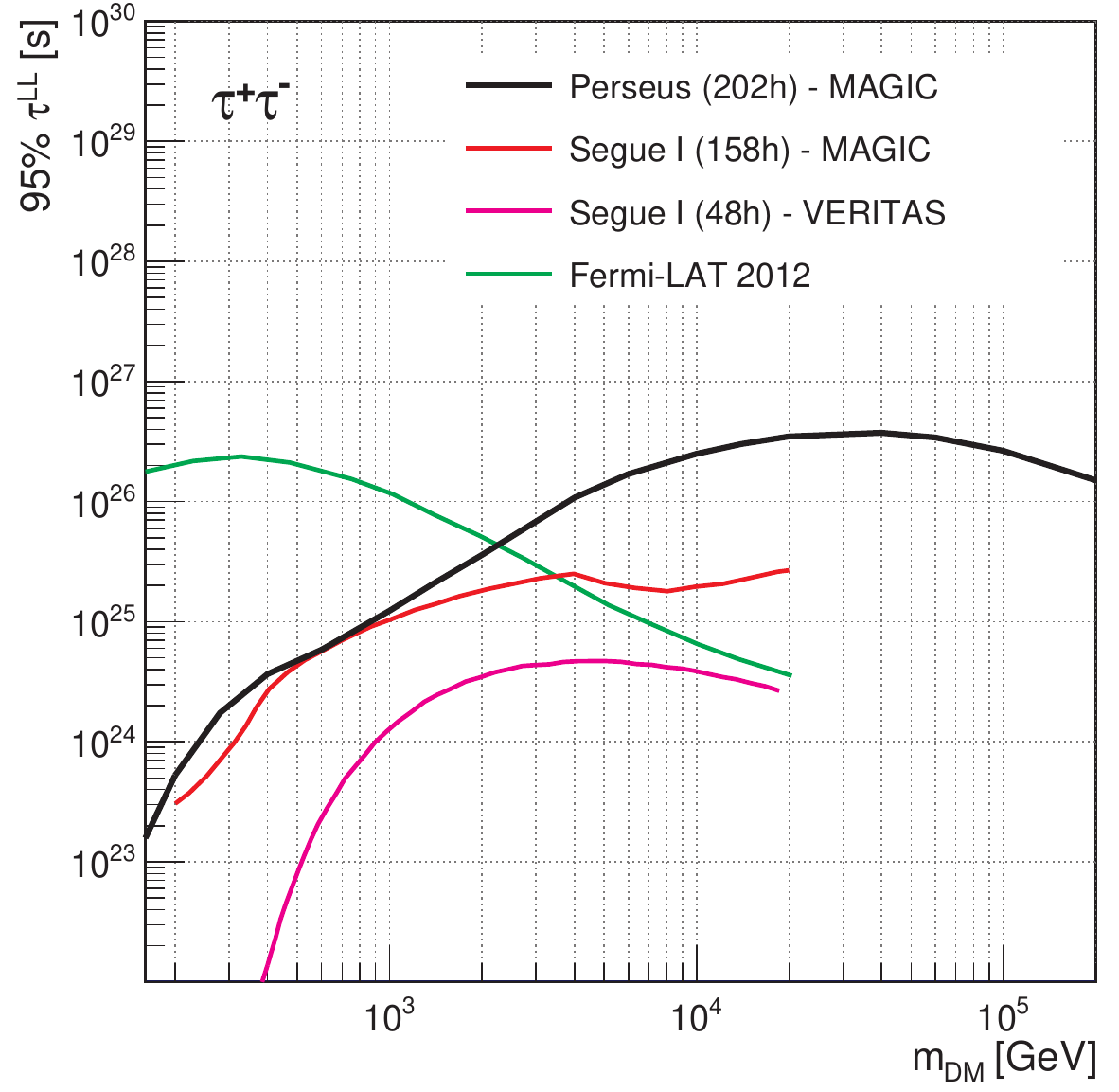}
   \includegraphics[width=.45\linewidth]{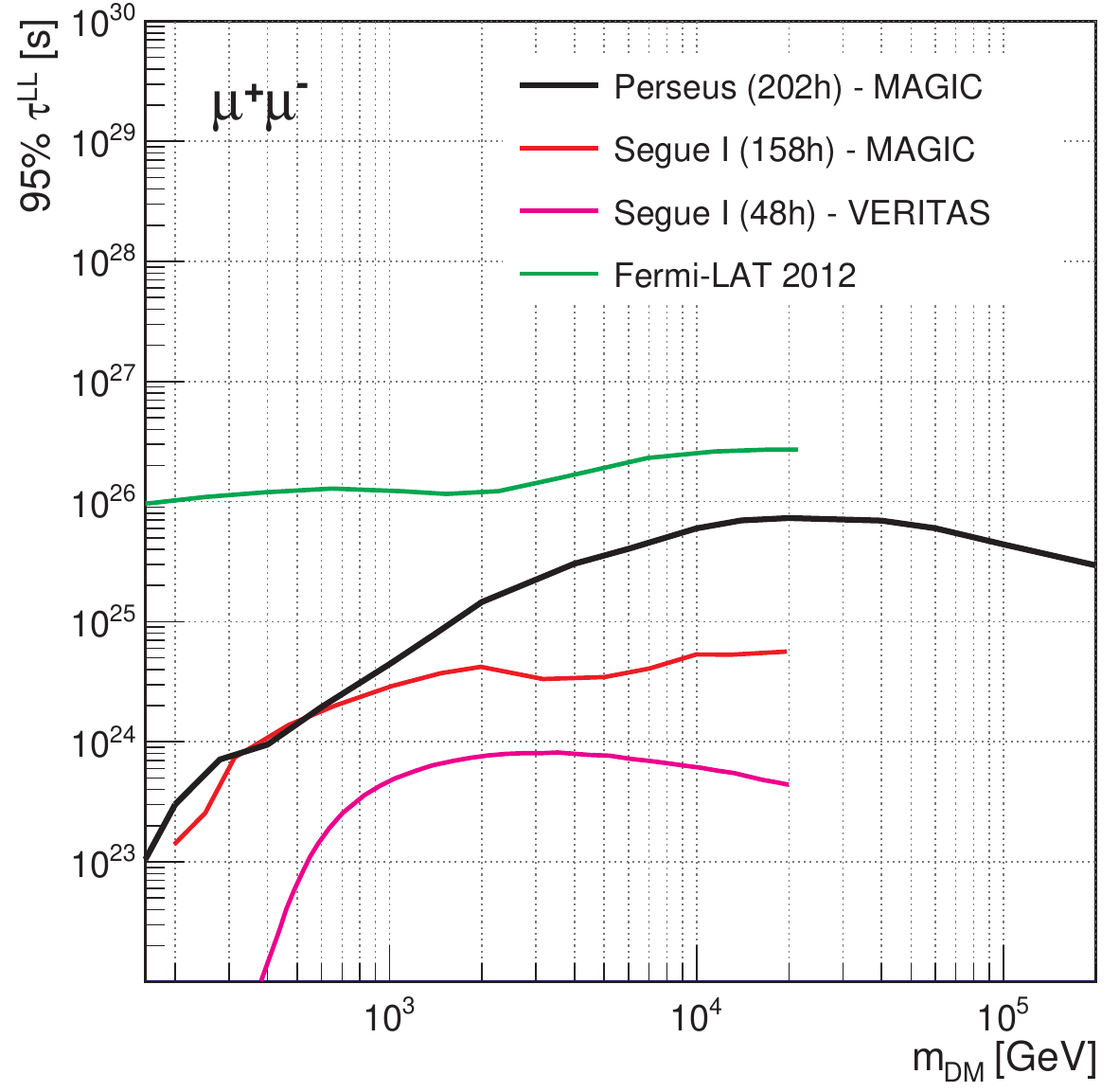}
\caption{\label{fig:compare_continuous}
    Comparison of the 95\% CL lower limit on the DM decay lifetime from the Perseus CG (solid black line) with
    similar measurements in the 
    %dwarf spheroidal galaxy 
    dSph Segue~I by the MAGIC~\citep[][red line]{Aleksic:2013xea} and 
    VERITAS~\citep[][pink line]{Aliu:2012ga} collaborations.
    The limits obtained from the diffuse galactic center from the Fermi-LAT collaboration~\citep[][green-line]{Ackermann:2012rg}
    are also shown (limits for $W^{+}W^{-}$ not available).}
\end{figure}
\begin{figure}[ht!]
  \centering
    \begin{subfigure}[b]{0.49\textwidth}
        \includegraphics[width=.9\linewidth]{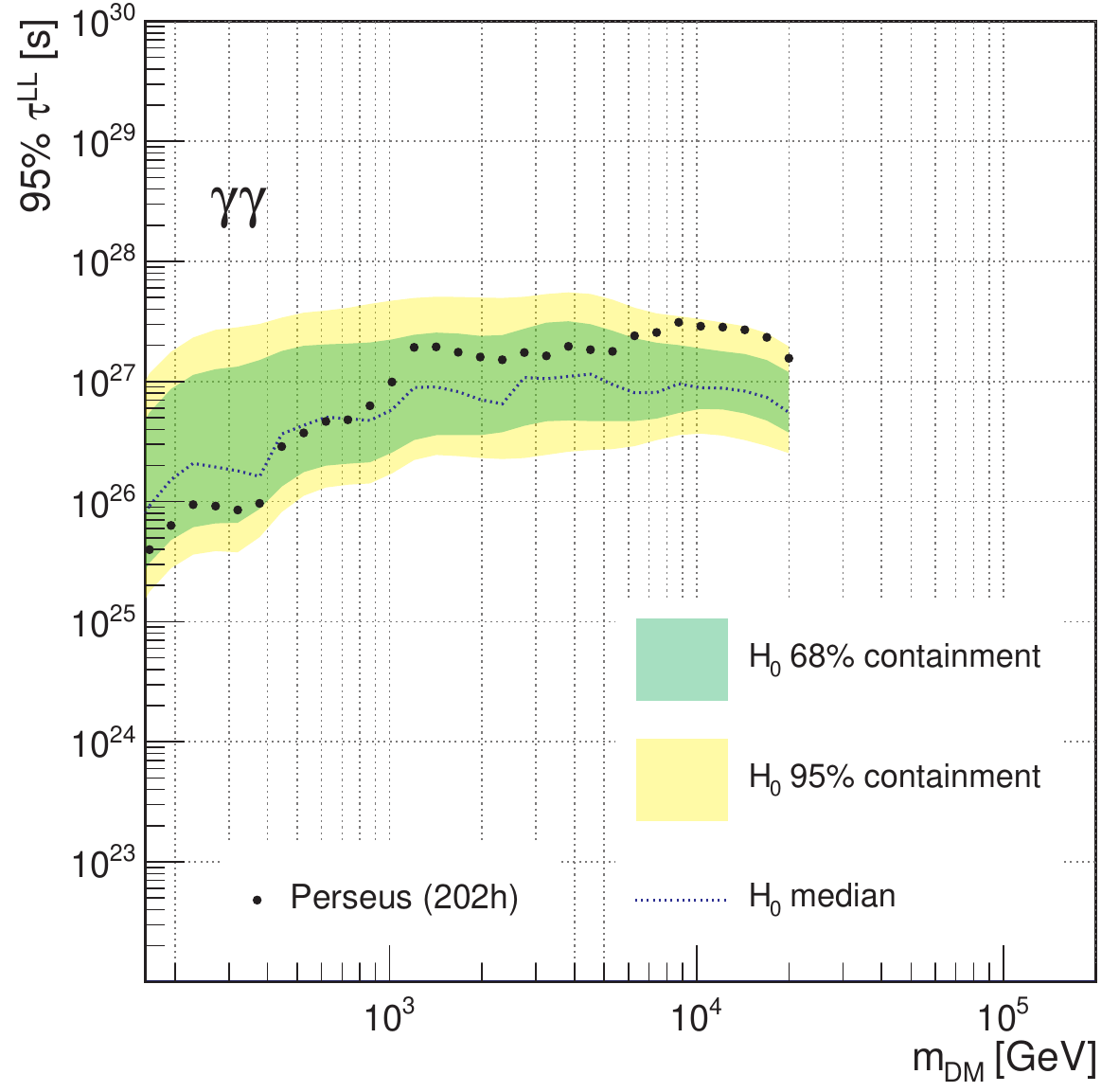}
        \caption{}
        \label{fig:results_lines}
    \end{subfigure}
    \begin{subfigure}[b]{0.49\textwidth}
         \includegraphics[width=.9\linewidth]{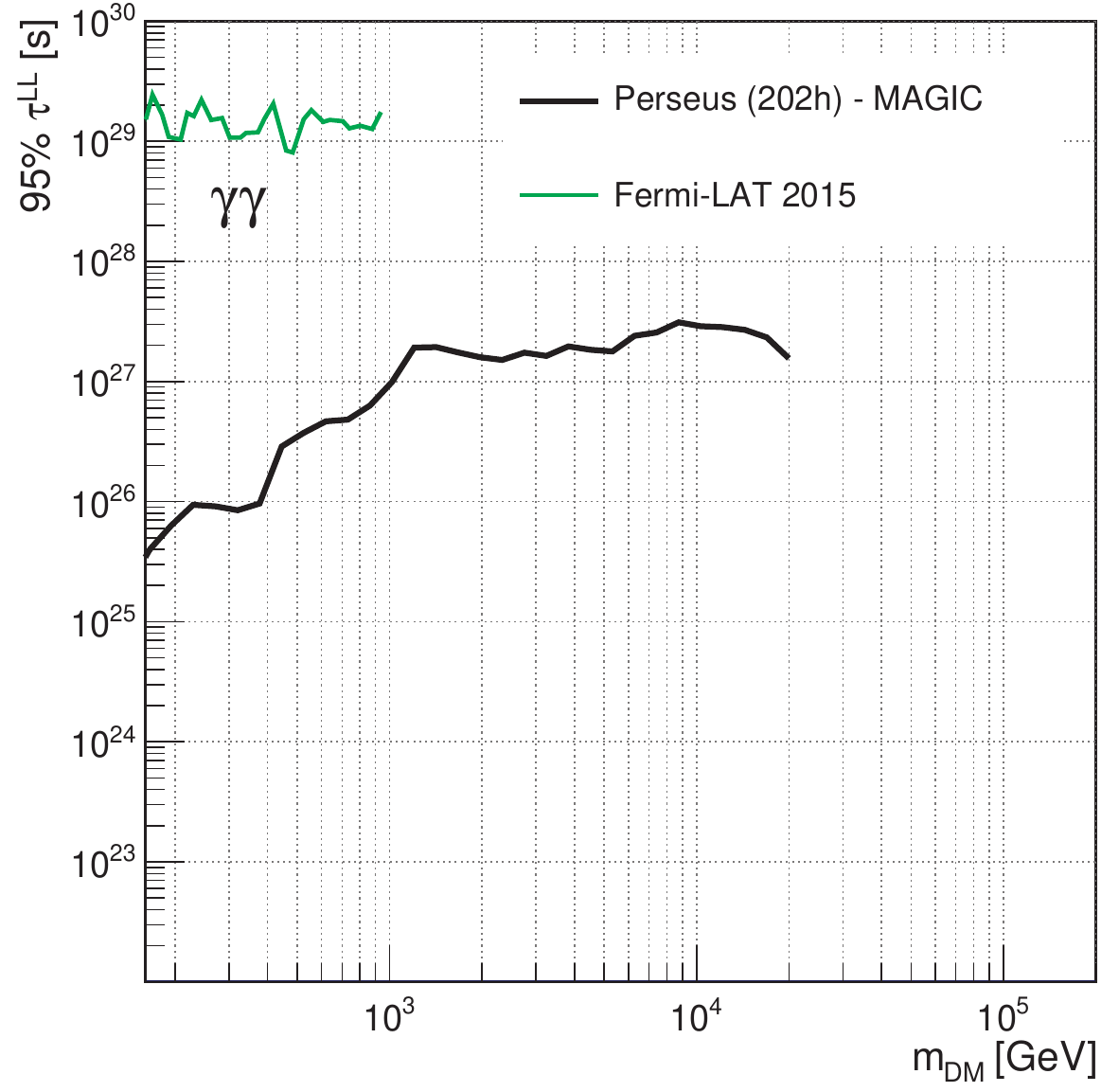}
        \caption{}
        \label{fig:compareresults_lines}
    \end{subfigure}
\caption{
(left) Same as~\autoref{fig:results_continuous} for DM particles decaying into $\gamma\gamma$.
The scanned mass range goes between 200 GeV and 20 TeV, 
since for larger masses the number of expected gamma-ray events detected by MAGIC in the observation time is lower than 1.
(right) Same as~\autoref{fig:compare_continuous} DM particles decaying into $\gamma\gamma$. Fermi-LAT data based on~\citep{Ackermann:2015lka}.}
\end{figure}
% \begin{figure}[ht!]
% \centering
%   \includegraphics[width=.45\linewidth]{figures/MAGIC_Paper_Perseus_Comparison_gammagamma.pdf}
% \caption{\label{fig:compareresults_lines}
%     Black solid (dashed) line shows the 95\% CL lower limit on decay life-time for  DM decaying into $\gamma\gamma$
%     for  DM masses from 200~\mbox{GeV} to 200~\mbox{TeV}
%     from 202~h of good quality data from the Perseus cluster of galaxy where
%     \gls{jfactor} is a fixed (nuisance) parameter. The XX line show the 95\% CL lower limits obtained with Fermi-LAT in \citep{Ackermann:2015lka} in the region of interest R180}
% \end{figure}

% In this section we present the results on the lifetime ($\tdm$) of  CG particles decaying into different SM particles achieved with selected $202$~h of good quality data  on the Perseus CG.
% The search is performed for  CG particles of masses between 200~$\GeV$ and 200~$\TeV$ for decays 
% into $\bb$, $\tautau$, $\mumu$ and $\WW$. 
% We also report on D CG particles decaying into $\gamma\gamma$ for masses between 
% 200~$\GeV$ to 20~$\GeV$,
% a decay mode largely discussed in the literature.
We performed a search for decaying DM in the Perseus CG using 202 h of data passing a thorough selection as described in~\autoref{sec:data}, 
for DM particles decaying into $\bb$, $\tautau$, $\mumu$, $\WW$, and $\gamma\gamma$, with masses between 200~GeV and 200~TeV
(for decays into $\gamma\gamma$, the scanned mass range has been reduced between 200~GeV and 20~TeV, 
since for such spectra for larger masses the number of expected gamma-ray events detected by MAGIC in the observation time is lower than 1).
95\% CL lower limits on the DM particle decay lifetime $\tauDM$ for each decay channel are obtained with a binned likelihood analysis 
(80 GeV to 10 TeV in 10 logarithmic-spaced bins\footnote{Empty bins are merged with neighbouring ones.}) 
%of the $202$~h Perseus data sample and 
using $J_{\text{dec}}=1.5\times10^{19}$ GeV cm$^{-2}$ (see~\autoref{sec:dm}).
The results for leptonic and hadronic decays are shown in ~\autoref{fig:results_continuous}  
%and \autoref{fig:results_lines} for monochromatic lines.
%The figures 
where also reported are the two-sided 68\% and 95\% containment bands and the median for the null hypothesis, 
computed from the distribution of the lower limits obtained from the analysis of 300
realizations of the null hypothesis. This consist of  MC simulations in which both ON and
OFF regions are generated from pure background probability density functions, 
assuming both similar exposures for the real data, 
and $\kappa_{i}$ taken as a nuisance parameter in the likelihood function.
We reach sensitivities $\tauDM>10^{26}$~s %for leptonic/hadronic channels 
%and $\tauDM>10^{27}$~s for $\gamma\gamma$ decays.
where no evidence for decaying DM is found in either decay mode.
\bigskip

\autoref{fig:compare_continuous} 
%and \autoref{fig:compareresults_lines} 
show the comparison of MAGIC lower limits
%in hadronic/leptonic decay lifetimes of decaying DM 
with results from other searches. Decaying DM scenarios are currently investigated with several classes of instruments and for different mass ranges. 
In the GeV-TeV mass range, 
%there is plenty of literature with constraints on decaying DM. 
the majority of limits make use of the Fermi-LAT instrument, sensitive in the MeV-GeV range, in two ways: either combining results from observations of CGs~\citep{Dugger:2010ys,Ke:2011xw,Zimmer:2011vy,Huang:2011xr,Charles:2016pgz,Ackermann:2015fdi} 
or making use of the integrated cosmological decaying DM contribution to the extragalactic diffuse light~\citep[see, e.g.,][]{Cirelli:2012ut,Cohen:2016uyg}, composed of prompt and secondary emission. 
It must be underlined that these results were not independently validated by the Fermi-LAT collaboration and that,
in some cases these limits depend on the model-dependent secondary components. 
In the TeV energy range, where searches for diffuse emission are hindered because of the limited FoV of ground based IACTs, the decaying DM case was discussed by \citet{Cirelli:2012ut}
%\footnote{H.E.S.S. data are not reported in \autoref{fig:compare_continuous} because they are not official collaboration results. In addition, the results in \citet{Cirelli:2012ut} concerning H.E.S.S. data were not updated after an erratum~\citep{Abramowski:2012au}.} 
showing lower limits on the DM decay lifetime with H.E.S.S. data for the Fornax CG,
but again an independent validation from the H.E.S.S. collaboration has not yet been published. 
At higher energies, the most stringent constraints on certain channels can be obtained with neutrinos with IceCube~\citep{Cohen:2016uyg} or ultra-high-energy cosmic rays with the Pierre Auger Observatory, KASKADE, and CASA-MIA~\citep{Aab:2015bza,Kang:2015gpa,Chantell:1997gs}. 
In the VHE gamma-ray range, our results are compared with previous limits obtained with MAGIC using 158~h of the dSph Segue~1~\citep{Aleksic:2013xea}. 
We also show limits from 48~h observation of Segue~1 with VERITAS~\citep{Aliu:2012ga} and with Fermi-LAT data on the Galactic Center~\citep{Ackermann:2012rg}. 
The comparison of these results may suffer from the fact that nuisance parameters are treated differently in different analyses.
Fermi-LAT results are more constraining in the low WIMP mass range, up to few hundreds of GeV, depending on the decay channel. However, 
they are based on significant assumptions on the nature of the diffuse galactic gamma-ray emission, as discussed in \cite{Ackermann:2012rg}.
The Perseus results are more constraining than previous %dwarf spheroidal galaxy 
dSph limits achieved above few hundreds
GeV and extend previous MAGIC results towards larger WIMP DM masses so far unexplored. 
From a mere comparison of the astrophysical factors (60 times larger for Perseus than for Segue~1)
%$J_{\text{dec}}$ ($1.5\times10^{19}~\mbox{GeV}\,\mbox{cm}^{-2}$ for Perseus and $\sim2.5\times 10^{17}~\mbox{GeV}\,\mbox{cm}^{-2}$ for Segue~1) 
one would have expected a similarly stronger constraint. However, several factors degrade the sensitivity in this analysis. Primarily, 
the region of interest is more complex: the presence of known astrophysical emitters in the FoV requires a reduction of the search signal region, 
and the extension of the DM profile induces a leakage of signal into the OFF region. 
%, but it has to be also taken into account that the sensitivities of the two analyses do not only
%depend on the $J$-factor, but also on:
%\begin{itemize}
    %\item[-] 
    %
    %\item[-] 
Secondly, this analysis is made more accurate by  additionally taking into account the uncertainty in the background rate estimation, 
and the different computation of the IRFs, which consider  the morphology of the CG emission. 
    %The fraction of events that are expected within our ROI ($\theta_{\text{min}}$ and $\theta_{\text{max}}$) is taken into account in the IRF (in particular the A$_{\text{eff}}$ in \autoref{eq:freeparameter}).
%\end{itemize}
%We underline how the treatment of systematics is different between the MAGIC CG and dSph analyses,
%namely the number of nuisances considered in this work has increased with respect our previous analysis.
%This partially explains the limited improvement with respect to the one expected by comparing the corresponding $J$-factors 
%(there are also the effect of the contamination of DM events in the OFF region that reduces even more the
%sensitivity of the analysis, or the larger signal integration angle, which increases also the integrated background).
%Indeed, a qualitative comparison between the two analysis by comparing the two $J$-factors may be miss-leading since
%the treatment of the $J_{\text{dec}}$-factor within the two analyses has also changed.
%In this work we have mentioned how, for Perseus, we consider a total $J_{\text{dec}}$-factor of $1.5\times10^{19}~\mbox{GeV}\,\mbox{cm}^{-2}$, that has been integrated for all the CG extension
%(and not between $\theta_{\text{min}}$ and $\theta_{\text{max}}$ as for the analysis of Segue~1).
%The fraction of events that are expected within our ROI is taken into account in the IRF 
%(in particular the A$_{\text_{eff}}$, that has been computed with the donut procedure, taking into account the whole CG extension).
\bigskip 

We did not consider the effect of a second gamma-ray contribution %to the continuous gamma-ray flux can 
coming from the interaction of charged particles
(most notably light leptons, generated during the decay process normally after hadronization, fragmentation and decay of prompt products) interacting with the 
intra-cluster magnetic field (ICMF) of Perseus or the cosmic microwave background (CMB).
These charged particles may diffuse away from the DM halo although they are expected to be contained by efficient energy loss such as with
synchrotron and Inverse Compton (IC) processes. 
This was investigated for the annihilating DM case in~\cite{Gomez-Vargas:2013bea} and \cite{Ackermann:2015fdi}, 
where the authors showed that for DM masses above 50 GeV, the contribution from secondary gamma rays can boost the signal up to a factor of 5 for muons and a factor of 2 for taus~\citep[Figure 5 of][]{Gomez-Vargas:2013bea} for the Galactic Center region.
Such a contribution would be proportional to the $J$-factor, and therefore be present also in our case.
The predominance of the former or the latter, discussed also in~\cite{Gomez-Vargas:2013bea}, is governed by 
the intensity of the ICMF. 
For values larger than $3\muG$, the magnetic field energy density is comparable to that of the CMB photons, 
that are the seeds for IC up-scatterings. 
Considering that the Perseus core is expected to have a magnetic field in the range $3-25\muG$ or larger 
\citep{Aleksic:2011cp,Taylor:2006ta}, the IC contribution is expected to be less relevant due to important synchrotron losses.
Above several hundreds GeV, our results are therefore conservative and could be only slightly more stringent in case
secondary emission is considered.
\newline

Finally, the results for monochromatic line decays are shown in ~\autoref{fig:results_lines}  
%and \autoref{fig:results_lines} for monochromatic lines.
where we reach sensitivities $\tauDM>10^{27}$~s.
%and $\tauDM>10^{27}$~s for $\gamma\gamma$ decays.
Again, no evidence for decaying DM is found.
In \autoref{fig:compareresults_lines}, Perseus results for line-like spectra are put into context and 
compared with the Fermi-LAT data of \citet{Ackermann:2015lka}\footnote{Using the lower limit on decay lifetime 
computed in the region of interest dubbed R180 in their paper, optimized to search for spectral lines from DM 
decay}. 
The Fermi-LAT collaboration has published several studies on spectral line searches~\citep{Abdo:2010nc,Ackermann:2012qk,Ackermann:2013uma}.  
In their latest work~\citep{Ackermann:2015lka}, Fermi-LAT updated their results using 5.8 years of Pass 8 data and
an optimized region of interest according to different DM realizations. 
One can see that Fermi-LAT data are more constraining below the TeV mass scale while Perseus results are the most
constraining results at the low TeV mass range and extend the current scanned mass range of decaying WIMPs to
larger values.

\section{Summary and Conclusions}
\label{sec:conclusions}
Clusters of galaxies are optimal targets for decaying DM searches, given their huge expected DM content. 
The MAGIC telescopes have observed the Perseus CG for about 400~h over several years. 
The data sample was used in this paper to search for decaying DM in the Perseus CG core. 
%The complexity of the \textbf{analysis has been increased} 
The analysis is made more complex
by the presence of the central radio galaxy NGC~1275, a known 
bright gamma-ray emitter with variable flux, as well as by the more peripheral radio galaxy IC~310. 
Furthermore, the putative signal region extends outwards to a radius a few times the telescope angular resolution. 
This necessitated the development of a tailored MC sample and 
the inclusion of extra terms in the likelihood function with respect to a standard analysis.
Out of the full data sample, 202~h have been selected for our study. 
No evidence of a DM signal has been found.
%but data were matching the null hypothesis in any case, in which no DM was presented. 
From this result, we cast lower limits on the decay lifetime $\tdm$ of WIMP DM
with masses between 200~GeV and 200~TeV for several ``pure'' (i.e. 100\% branching ratio) 
decay channels $b\bar{b},\; \tau^{+}\tau^{-},\; W^{+}W^{-},\; \mu^{+}\mu^{-}$
%(see \autoref{fig:results_continuous}) 
as well as for ``pure'' $\gamma\gamma$ decay line. % (see \autoref{fig:results_lines}).
We have reached strong limits on the order of $\tdm=2\times10^{26}$~s for a 10 TeV DM particle decaying into $\tau^{+}\tau^{-}$ 
and $\tdm=3\times10^{27}$~s for a 10 TeV DM particle decaying into $\gamma\gamma$.
Our limits improve previous MAGIC results and are the most constraining limits on the decay lifetime
of DM particles based on observations from ground-based gamma-ray instruments.
\bigskip

It is unlikely that dwarf satellite galaxies can provide stronger constraints on decaying DM 
scenarios than the ones provided by CGs. Improving current limits on decaying DM particles with 
this technique would require a significant increase of observation time on 
CGs, which is unlikely in the close future with the current generation of IACT. 
Therefore, these results are unlikely to be further improved until the next generation of Cherenkov telescopes, like the Cherenkov Telescope Array~\citep[CTA,][]{Acharya:2017ttl}, becomes active.

\paragraph{Acknowledgements}
We would like to thank the Instituto de Astrof\'{\i}sica de Canarias for the excellent working conditions at the Observatorio del Roque de los Muchachos in La Palma. The financial support of the German BMBF and MPG, the Italian INFN and INAF, the Swiss National Fund SNF, the ERDF under the Spanish MINECO (FPA2015-69818-P, FPA2012-36668, FPA2015-68378-P, FPA2015-69210-C6-2-R, FPA2015-69210-C6-4-R, FPA2015-69210-C6-6-R, AYA2015-71042-P, AYA2016-76012-C3-1-P, ESP2015-71662-C2-2-P, CSD2009-00064), and the Japanese JSPS and MEXT is gratefully acknowledged. This work was also supported by the Spanish Centro de Excelencia ``Severo Ochoa'' SEV-2012-0234 and SEV-2015-0548, and Unidad de Excelencia ``Mar\'{\i}a de Maeztu'' MDM-2014-0369, by the Croatian Science Foundation (HrZZ) Project IP-2016-06-9782 and the University of Rijeka Project 13.12.1.3.02, by the DFG Collaborative Research Centers SFB823/C4 and SFB876/C3, the Polish National Research Centre grant UMO-2016/22/M/ST9/00382 and by the Brazilian MCTIC, CNPq and FAPERJ.
The work of the author M. Vazquez Acosta is financed with grant RYC-2013-14660 of MINECO.
\newline 

We are also indepted to G. Brunetti, C. Combet, D. Maurin, S. Zimmer for
valid inputs and specially M.A. Sanchez Conde and F. Zandanel, for fruitful  discussions and insights. 
\bibliographystyle{abbrvnat}
\bibliography{biblio}

\begin{thebibliography}{74}
\providecommand{\natexlab}[1]{#1}
\providecommand{\url}[1]{\texttt{#1}}
\expandafter\ifx\csname urlstyle\endcsname\relax
  \providecommand{\doi}[1]{doi: #1}\else
  \providecommand{\doi}{doi: \begingroup \urlstyle{rm}\Url}\fi

\bibitem[Aab et~al.(2015)]{Aab:2015bza}
A.~Aab et~al.
\newblock {The Pierre Auger Observatory: Contributions to the 34th
  International Cosmic Ray Conference (ICRC 2015)}.
\newblock 2015.
\newblock URL
  \url{https://inspirehep.net/record/1393211/files/arXiv:1509.03732.pdf}.

\bibitem[Abdo et~al.(2009)]{Abdo:2009zk}
A.~A. Abdo et~al.
\newblock {Measurement of the Cosmic Ray e+ plus e- spectrum from 20 GeV to 1
  TeV with the Fermi Large Area Telescope}.
\newblock \emph{Phys. Rev. Lett.}, 102:\penalty0 181101, 2009.
\newblock \doi{10.1103/PhysRevLett.102.181101}.

\bibitem[Abdo et~al.(2010)]{Abdo:2010nc}
A.~A. Abdo et~al.
\newblock {Fermi LAT Search for Photon Lines from 30 to 200 GeV and Dark Matter
  Implications}.
\newblock \emph{Phys. Rev. Lett.}, 104:\penalty0 091302, 2010.
\newblock \doi{10.1103/PhysRevLett.104.091302}.

\bibitem[Acharya et~al.(2017)]{Acharya:2017ttl}
B.~S. Acharya et~al.
\newblock {Science with the Cherenkov Telescope Array}.
\newblock 2017.

\bibitem[Ackermann et~al.(2010)]{Ackermann:2010ij}
M.~Ackermann et~al.
\newblock {Fermi LAT observations of cosmic-ray electrons from 7 GeV to 1 TeV}.
\newblock \emph{Phys. Rev.}, D82:\penalty0 092004, 2010.
\newblock \doi{10.1103/PhysRevD.82.092004}.

\bibitem[Ackermann et~al.(2012{\natexlab{a}})]{Ackermann:2012qk}
M.~Ackermann et~al.
\newblock {Fermi LAT Search for Dark Matter in Gamma-ray Lines and the
  Inclusive Photon Spectrum}.
\newblock \emph{Phys. Rev.}, D86:\penalty0 022002, 2012{\natexlab{a}}.
\newblock \doi{10.1103/PhysRevD.86.022002}.

\bibitem[Ackermann et~al.(2012{\natexlab{b}})]{Ackermann:2012rg}
M.~Ackermann et~al.
\newblock {Constraints on the Galactic Halo Dark Matter from Fermi-LAT Diffuse
  Measurements}.
\newblock \emph{Astrophys. J.}, 761:\penalty0 91, 2012{\natexlab{b}}.
\newblock \doi{10.1088/0004-637X/761/2/91}.

\bibitem[Ackermann et~al.(2013)]{Ackermann:2013uma}
M.~Ackermann et~al.
\newblock {Search for Gamma-ray Spectral Lines with the Fermi Large Area
  Telescope and Dark Matter Implications}.
\newblock \emph{Phys. Rev.}, D88:\penalty0 082002, 2013.
\newblock \doi{10.1103/PhysRevD.88.082002}.

\bibitem[Ackermann et~al.(2015{\natexlab{a}})]{Ackermann:2015fdi}
M.~Ackermann et~al.
\newblock {Search for extended gamma-ray emission from the Virgo galaxy cluster
  with Fermi-LAT}.
\newblock \emph{Astrophys. J.}, 812\penalty0 (2):\penalty0 159,
  2015{\natexlab{a}}.
\newblock \doi{10.1088/0004-637X/812/2/159}.

\bibitem[Ackermann et~al.(2015{\natexlab{b}})]{Ackermann:2015lka}
M.~Ackermann et~al.
\newblock {Updated search for spectral lines from Galactic dark matter
  interactions with pass 8 data from the Fermi Large Area Telescope}.
\newblock \emph{Phys. Rev.}, D91\penalty0 (12):\penalty0 122002,
  2015{\natexlab{b}}.
\newblock \doi{10.1103/PhysRevD.91.122002}.

\bibitem[Adriani et~al.(2009)]{Adriani:2008zr}
O.~Adriani et~al.
\newblock {An anomalous positron abundance in cosmic rays with energies 1.5-100
  GeV}.
\newblock \emph{Nature}, 458:\penalty0 607--609, 2009.
\newblock \doi{10.1038/nature07942}.

\bibitem[Aguilar et~al.(2013)]{Aguilar:2013qda}
M.~Aguilar et~al.
\newblock {First Result from the Alpha Magnetic Spectrometer on the
  International Space Station: Precision Measurement of the Positron Fraction
  in Primary Cosmic Rays of 0.5–350 GeV}.
\newblock \emph{Phys. Rev. Lett.}, 110:\penalty0 141102, 2013.
\newblock \doi{10.1103/PhysRevLett.110.141102}.

\bibitem[Ahnen et~al.(2016{\natexlab{a}})]{Ahnen:2016qkt}
M.~L. Ahnen et~al.
\newblock {Deep observation of the NGC 1275 region with MAGIC: search of
  diffuse $\gamma$-ray emission from cosmic rays in the Perseus cluster}.
\newblock \emph{Astron. Astrophys.}, 589:\penalty0 A33, 2016{\natexlab{a}}.
\newblock \doi{10.1051/0004-6361/201527846}.

\bibitem[Ahnen et~al.(2016{\natexlab{b}})]{Ahnen:2016qkx}
M.~L. Ahnen et~al.
\newblock {Limits to dark matter annihilation cross-section from a combined
  analysis of MAGIC and Fermi-LAT observations of dwarf satellite galaxies}.
\newblock \emph{JCAP}, 1602\penalty0 (02):\penalty0 039, 2016{\natexlab{b}}.
\newblock \doi{10.1088/1475-7516/2016/02/039}.

\bibitem[Ahnen et~al.(2017)]{Ahnen:2017vsf}
M.~L. Ahnen et~al.
\newblock {Performance of the MAGIC telescopes under moonlight}.
\newblock \emph{Astropart. Phys.}, 94:\penalty0 29--41, 2017.
\newblock \doi{10.1016/j.astropartphys.2017.08.001}.

\bibitem[Ahnen et~al.(2018)]{Ahnen:2017pqx}
M.~L. Ahnen et~al.
\newblock {Indirect dark matter searches in the dwarf satellite galaxy Ursa
  Major II with the MAGIC Telescopes}.
\newblock \emph{JCAP}, 1803\penalty0 (03):\penalty0 009, 2018.
\newblock \doi{10.1088/1475-7516/2018/03/009}.

\bibitem[Albert et~al.(2008)]{Albert:2007yd}
J.~Albert et~al.
\newblock {Implementation of the Random Forest Method for the Imaging
  Atmospheric Cherenkov Telescope MAGIC}.
\newblock \emph{Nucl. Instrum. Meth.}, A588:\penalty0 424--432, 2008.
\newblock \doi{10.1016/j.nima.2007.11.068}.

\bibitem[Aleksi\'c et~al.(2012{\natexlab{a}})Aleksi\'c, Rico, and
  Martinez]{Aleksic:2012cp}
J.~Aleksi\'c, J.~Rico, and M.~Martinez.
\newblock {Optimized analysis method for indirect dark matter searches with
  Imaging Air Cherenkov Telescopes}.
\newblock \emph{JCAP}, 1210:\penalty0 032, 2012{\natexlab{a}}.
\newblock \doi{10.1088/1475-7516/2012/10/032}.

\bibitem[Aleksi\'c et~al.(2010{\natexlab{a}})]{Aleksic:2009ir}
J.~Aleksi\'c et~al.
\newblock {MAGIC Gamma-Ray Telescope Observation of the Perseus Cluster of
  Galaxies: Implications for Cosmic Rays, Dark Matter and NGC 1275}.
\newblock \emph{Astrophys. J.}, 710:\penalty0 634--647, 2010{\natexlab{a}}.
\newblock \doi{10.1088/0004-637X/710/1/634}.

\bibitem[Aleksi\'c et~al.(2010{\natexlab{b}})]{Aleksic:2010xk}
J.~Aleksi\'c et~al.
\newblock {Detection of very high energy gamma-ray emission from the Perseus
  cluster head-tail galaxy IC 310 by the MAGIC telescopes}.
\newblock \emph{Astrophys. J.}, 723:\penalty0 L207, 2010{\natexlab{b}}.
\newblock \doi{10.1088/2041-8205/723/2/L207}.

\bibitem[Aleksi\'c et~al.(2012{\natexlab{b}})]{Aleksic:2011bx}
J.~Aleksi\'c et~al.
\newblock {Performance of the MAGIC stereo system obtained with Crab Nebula
  data}.
\newblock \emph{Astropart. Phys.}, 35:\penalty0 435--448, 2012{\natexlab{b}}.
\newblock \doi{10.1016/j.astropartphys.2011.11.007}.

\bibitem[Aleksi\'c et~al.(2012{\natexlab{c}})]{Aleksic:2011cp}
J.~Aleksi\'c et~al.
\newblock {Constraining Cosmic Rays and Magnetic Fields in the Perseus Galaxy
  Cluster with TeV observations by the MAGIC telescopes}.
\newblock \emph{Astron. Astrophys.}, 541:\penalty0 A99, 2012{\natexlab{c}}.
\newblock \doi{10.1051/0004-6361/201118502}.

\bibitem[Aleksi\'c et~al.(2012{\natexlab{d}})]{Aleksic:2011eb}
J.~Aleksi\'c et~al.
\newblock {Detection of very high energy gamma-ray emission from NGC 1275 by
  the MAGIC telescopes}.
\newblock \emph{Astron. Astrophys.}, 539:\penalty0 L2, 2012{\natexlab{d}}.
\newblock \doi{10.1051/0004-6361/201118668}.

\bibitem[Aleksi\'c et~al.(2014{\natexlab{a}})]{Aleksic:2013bya}
J.~Aleksi\'c et~al.
\newblock {Rapid and multiband variability of the TeV bright active nucleus of
  the galaxy IC 310}.
\newblock \emph{Astron. Astrophys.}, 563:\penalty0 A91, 2014{\natexlab{a}}.
\newblock \doi{10.1051/0004-6361/201321938}.

\bibitem[Aleksi\'c et~al.(2014{\natexlab{b}})]{Aleksic:2013kaa}
J.~Aleksi\'c et~al.
\newblock {Contemporaneous observations of the radio galaxy NGC 1275 from radio
  to very high energy $\gamma$-rays}.
\newblock \emph{Astron. Astrophys.}, 564:\penalty0 A5, 2014{\natexlab{b}}.
\newblock \doi{10.1051/0004-6361/201322951}.

\bibitem[Aleksi\'c et~al.(2014{\natexlab{c}})]{Aleksic:2013xea}
J.~Aleksi\'c et~al.
\newblock {Optimized dark matter searches in deep observations of Segue 1 with
  MAGIC}.
\newblock \emph{JCAP}, 1402:\penalty0 008, 2014{\natexlab{c}}.
\newblock \doi{10.1088/1475-7516/2014/02/008}.

\bibitem[Aleksi\'c et~al.(2014{\natexlab{d}})]{Aleksic:2014xsg}
J.~Aleksi\'c et~al.
\newblock {Black hole lightning due to particle acceleration at subhorizon
  scales}.
\newblock \emph{Science}, 346:\penalty0 1080--1084, 2014{\natexlab{d}}.
\newblock \doi{10.1126/science.1256183}.

\bibitem[Aleksi\'c et~al.(2016{\natexlab{a}})]{Aleksic:2014lkm}
J.~Aleksi\'c et~al.
\newblock {The major upgrade of the MAGIC telescopes, Part II: A performance
  study using observations of the Crab Nebula}.
\newblock \emph{Astropart. Phys.}, 72:\penalty0 76--94, 2016{\natexlab{a}}.
\newblock \doi{10.1016/j.astropartphys.2015.02.005}.

\bibitem[Aleksi\'c et~al.(2016{\natexlab{b}})]{Aleksic:2014poa}
J.~Aleksi\'c et~al.
\newblock {The major upgrade of the MAGIC telescopes, Part I: The hardware
  improvements and the commissioning of the system}.
\newblock \emph{Astropart. Phys.}, 72:\penalty0 61--75, 2016{\natexlab{b}}.
\newblock \doi{10.1016/j.astropartphys.2015.04.004}.

\bibitem[Aliu et~al.(2012)]{Aliu:2012ga}
E.~Aliu et~al.
\newblock {VERITAS Deep Observations of the Dwarf Spheroidal Galaxy Segue 1}.
\newblock \emph{Phys. Rev.}, D85:\penalty0 062001, 2012.
\newblock \doi{10.1103/PhysRevD.85.062001, 10.1103/PhysRevD.91.129903}.
\newblock [Erratum: Phys. Rev.D91,no.12,129903(2015)].

\bibitem[Ando and Ishiwata(2015)]{Ando:2015qda}
S.~Ando and K.~Ishiwata.
\newblock {Constraints on decaying dark matter from the extragalactic gamma-ray
  background}.
\newblock \emph{JCAP}, 1505\penalty0 (05):\penalty0 024, 2015.
\newblock \doi{10.1088/1475-7516/2015/05/024}.

\bibitem[Ando and Kusenko(2010)]{Ando:2010ye}
S.~Ando and A.~Kusenko.
\newblock {Interactions of keV sterile neutrinos with matter}.
\newblock \emph{Phys. Rev.}, D81:\penalty0 113006, 2010.
\newblock \doi{10.1103/PhysRevD.81.113006}.

\bibitem[Berezinsky et~al.(1991)Berezinsky, Masiero, and
  Valle]{Berezinsky:1991sp}
V.~Berezinsky, A.~Masiero, and J.~W.~F. Valle.
\newblock {Cosmological signatures of supersymmetry with spontaneously broken
  R-parity}.
\newblock \emph{Phys. Lett.}, B266:\penalty0 382--388, 1991.
\newblock \doi{10.1016/0370-2693(91)91055-Z}.

\bibitem[Bergstrom et~al.(1998)Bergstrom, Ullio, and Buckley]{Bergstrom:1997fj}
L.~Bergstrom, P.~Ullio, and J.~H. Buckley.
\newblock {Observability of gamma-rays from dark matter neutralino
  annihilations in the Milky Way halo}.
\newblock \emph{Astropart. Phys.}, 9:\penalty0 137--162, 1998.
\newblock \doi{10.1016/S0927-6505(98)00015-2}.

\bibitem[Boehm et~al.(2004)Boehm, Ensslin, and Silk]{Boehm:2002yz}
C.~Boehm, T.~A. Ensslin, and J.~Silk.
\newblock {Can Annihilating dark matter be lighter than a few GeVs?}
\newblock \emph{J. Phys.}, G30:\penalty0 279--286, 2004.
\newblock \doi{10.1088/0954-3899/30/3/004}.

\bibitem[Cat\'a et~al.(2017)Cat\'a, Ibarra, and Ingenhütt]{Cata:2016epa}
O.~Cat\'a, A.~Ibarra, and S.~Ingenhütt.
\newblock {Dark matter decay through gravity portals}.
\newblock \emph{Phys. Rev.}, D95\penalty0 (3):\penalty0 035011, 2017.
\newblock \doi{10.1103/PhysRevD.95.035011}.

\bibitem[Chantell et~al.(1997)]{Chantell:1997gs}
M.~C. Chantell et~al.
\newblock {Limits on the isotropic diffuse flux of ultrahigh-energy gamma
  radiation}.
\newblock \emph{Phys. Rev. Lett.}, 79:\penalty0 1805--1808, 1997.
\newblock \doi{10.1103/PhysRevLett.79.1805}.

\bibitem[Charles et~al.(2016)]{Charles:2016pgz}
E.~Charles et~al.
\newblock {Sensitivity Projections for Dark Matter Searches with the Fermi
  Large Area Telescope}.
\newblock \emph{Phys. Rept.}, 636:\penalty0 1--46, 2016.
\newblock \doi{10.1016/j.physrep.2016.05.001}.

\bibitem[Chen and Kamionkowski(2004)]{Chen:2003gz}
X.-L. Chen and M.~Kamionkowski.
\newblock {Particle decays during the cosmic dark ages}.
\newblock \emph{Phys. Rev.}, D70:\penalty0 043502, 2004.
\newblock \doi{10.1103/PhysRevD.70.043502}.

\bibitem[Chen et~al.(2007)Chen, Reiprich, Bohringer, Ikebe, and
  Zhang]{Chen:2007sz}
Y.~Chen, T.~H. Reiprich, H.~Bohringer, Y.~Ikebe, and Y.~Y. Zhang.
\newblock {Statistics of X-ray observables for the cooling-core and non-cooling
  core galaxy clusters}.
\newblock \emph{Astron. Astrophys.}, 2007.
\newblock \doi{10.1051/0004-6361:20066471}.
\newblock [Astron. Astrophys.466,805(2007)].

\bibitem[Cirelli et~al.(2012)Cirelli, Moulin, Panci, Serpico, and
  Viana]{Cirelli:2012ut}
M.~Cirelli, E.~Moulin, P.~Panci, P.~D. Serpico, and A.~Viana.
\newblock {Gamma ray constraints on Decaying Dark Matter}.
\newblock \emph{Phys. Rev.}, D86:\penalty0 083506, 2012.
\newblock \doi{10.1103/PhysRevD.86.083506, 10.1103/PhysRevD.86.109901}.

\bibitem[Cohen et~al.(2017)Cohen, Murase, Rodd, Safdi, and
  Soreq]{Cohen:2016uyg}
T.~Cohen, K.~Murase, N.~L. Rodd, B.~R. Safdi, and Y.~Soreq.
\newblock {$\gamma$-ray Constraints on Decaying Dark Matter and Implications
  for IceCube}.
\newblock \emph{Phys. Rev. Lett.}, 119\penalty0 (2):\penalty0 021102, 2017.
\newblock \doi{10.1103/PhysRevLett.119.021102}.

\bibitem[Dugger et~al.(2010)Dugger, Jeltema, and Profumo]{Dugger:2010ys}
L.~Dugger, T.~E. Jeltema, and S.~Profumo.
\newblock {Constraints on Decaying Dark Matter from Fermi Observations of
  Nearby Galaxies and Clusters}.
\newblock \emph{JCAP}, 1012:\penalty0 015, 2010.
\newblock \doi{10.1088/1475-7516/2010/12/015}.

\bibitem[Feng(2010)]{Feng:2010gw}
J.~L. Feng.
\newblock {Dark Matter Candidates from Particle Physics and Methods of
  Detection}.
\newblock \emph{Ann. Rev. Astron. Astrophys.}, 48:\penalty0 495--545, 2010.
\newblock \doi{10.1146/annurev-astro-082708-101659}.

\bibitem[Feng et~al.(2014)Feng, Yang, He, Dong, Fan, and Chang]{Feng:2014zca}
L.~Feng, R.-Z. Yang, H.-N. He, T.-K. Dong, Y.-Z. Fan, and J.~Chang.
\newblock {AMS-02 positron excess: new bounds on dark matter models and hint
  for primary electron spectrum hardening}.
\newblock \emph{Phys. Lett.}, B728:\penalty0 250--255, 2014.
\newblock \doi{10.1016/j.physletb.2013.12.012}.

\bibitem[Fomin et~al.(1994)Fomin, Stepanian, Lamb, Lewis, Punch, and
  Weekes]{Fomin:1994aj}
V.~P. Fomin, A.~A. Stepanian, R.~C. Lamb, D.~A. Lewis, M.~Punch, and T.~C.
  Weekes.
\newblock {New methods of atmospheric Cherenkov imaging for gamma-ray
  astronomy. 1: The False source method}.
\newblock \emph{Astropart. Phys.}, 2:\penalty0 137--150, 1994.
\newblock \doi{10.1016/0927-6505(94)90036-1}.

\bibitem[Freese(2009)]{Freese:2008cz}
K.~Freese.
\newblock {Review of Observational Evidence for Dark Matter in the Universe and
  in upcoming searches for Dark Stars}.
\newblock \emph{EAS Publ. Ser.}, 36:\penalty0 113--126, 2009.
\newblock \doi{10.1051/eas/0936016}.

\bibitem[Fruck et~al.(2014)Fruck, Gaug, Zanin, Dorner, Garrido, Mirzoyan, and
  Font]{Fruck:2014mja}
C.~Fruck, M.~Gaug, R.~Zanin, D.~Dorner, D.~Garrido, R.~Mirzoyan, and L.~Font.
\newblock {A novel LIDAR-based Atmospheric Calibration Method for Improving the
  Data Analysis of MAGIC}.
\newblock In \emph{{Proceedings, 33rd International Cosmic Ray Conference
  (ICRC2013): Rio de Janeiro, Brazil, July 2-9, 2013}}, page 1054, 2014.
\newblock URL
  \url{http://inspirehep.net/record/1285998/files/arXiv:1403.3591.pdf}.

\bibitem[Garny et~al.(2011)Garny, Ibarra, Tran, and Weniger]{Garny:2011}
M.~Garny, A.~Ibarra, D.~Tran, and C.~Weniger.
\newblock {Gamma-ray lines from radiative dark matter decay}.
\newblock \emph{Journal of Cosmology and Astroparticle Physics}, 2011\penalty0
  (01):\penalty0 032--032, 2011.
\newblock ISSN 1475-7516.
\newblock \doi{10.1088/1475-7516/2011/01/032}.
\newblock URL
  \url{http://stacks.iop.org/1475-7516/2011/i=01/a=032?key=crossref.a672f146155cad875c8a0c2c07f155c0}.

\bibitem[G\'omez-Vargas et~al.(2013)G\'omez-Vargas, S\'anchez-Conde, Huh,
  Peir\'o, Prada, Morselli, Klypin, Cerdeño, Mambrini, and
  Muñoz]{Gomez-Vargas:2013bea}
G.~A. G\'omez-Vargas, M.~A. S\'anchez-Conde, J.-H. Huh, M.~Peir\'o, F.~Prada,
  A.~Morselli, A.~Klypin, D.~G. Cerdeño, Y.~Mambrini, and C.~Muñoz.
\newblock {Constraints on WIMP annihilation for contracted dark matterin the
  inner Galaxy with the Fermi-LAT}.
\newblock \emph{JCAP}, 1310:\penalty0 029, 2013.
\newblock \doi{10.1088/1475-7516/2013/10/029}.

\bibitem[Griest and Kamionkowski(1990)]{Griest:1989wd}
K.~Griest and M.~Kamionkowski.
\newblock {Unitarity Limits on the Mass and Radius of Dark Matter Particles}.
\newblock \emph{Phys. Rev. Lett.}, 64:\penalty0 615, 1990.
\newblock \doi{10.1103/PhysRevLett.64.615}.

\bibitem[Hernquist(1990)]{Hernquist:1990be}
L.~Hernquist.
\newblock {An Analytical Model for Spherical Galaxies and Bulges}.
\newblock \emph{Astrophys. J.}, 356:\penalty0 359, 1990.
\newblock \doi{10.1086/168845}.

\bibitem[Huang et~al.(2012)Huang, Vertongen, and Weniger]{Huang:2011xr}
X.~Huang, G.~Vertongen, and C.~Weniger.
\newblock {Probing Dark Matter Decay and Annihilation with Fermi LAT
  Observations of Nearby Galaxy Clusters}.
\newblock \emph{JCAP}, 1201:\penalty0 042, 2012.
\newblock \doi{10.1088/1475-7516/2012/01/042}.

\bibitem[Ibarra(2012)]{Ibarra:2012a}
A.~Ibarra.
\newblock {Gamma-ray lines from dark matter decay}.
\newblock \emph{Journal of physics Conference Series}, 384:\penalty0 012001,
  2012.
\newblock \doi{doi:10.1088/1742-6596/384/1/012001}.
\newblock URL
  \url{http://iopscience.iop.org/article/10.1088/1742-6596/384/1/012001/pdf}.

\bibitem[Jeltema et~al.(2009)Jeltema, Kehayias, and Profumo]{Jeltema:2008vu}
T.~E. Jeltema, J.~Kehayias, and S.~Profumo.
\newblock {Gamma Rays from Clusters and Groups of Galaxies: Cosmic Rays versus
  Dark Matter}.
\newblock \emph{Phys. Rev.}, D80:\penalty0 023005, 2009.
\newblock \doi{10.1103/PhysRevD.80.023005}.

\bibitem[Kang et~al.(2015)]{Kang:2015gpa}
D.~Kang et~al.
\newblock {A limit on the diffuse gamma-rays measured with KASCADE-Grande}.
\newblock \emph{J. Phys. Conf. Ser.}, 632\penalty0 (1):\penalty0 012013, 2015.
\newblock \doi{10.1088/1742-6596/632/1/012013}.

\bibitem[Ke et~al.(2011)Ke, Luo, Wang, and Zhu]{Ke:2011xw}
J.~Ke, M.~Luo, L.~Wang, and G.~Zhu.
\newblock {Gamma-rays from Nearby Clusters: Constraints on Selected Decaying
  Dark Matter Models}.
\newblock \emph{Phys. Lett.}, B698:\penalty0 44--51, 2011.
\newblock \doi{10.1016/j.physletb.2011.02.055}.

\bibitem[{Li} and {Ma}(1983)]{LiMa:1983ApJ}
T.-P. {Li} and Y.-Q. {Ma}.
\newblock {Analysis methods for results in gamma-ray astronomy}.
\newblock \emph{\apj}, 272:\penalty0 317--324, Sept. 1983.
\newblock \doi{10.1086/161295}.

\bibitem[Molin\'e et~al.(2017)Molin\'e, S\'anchez-Conde, Palomares-Ruiz, and
  Prada]{Moline:2016pbm}
A.~Molin\'e, M.~A. S\'anchez-Conde, S.~Palomares-Ruiz, and F.~Prada.
\newblock {Characterization of subhalo structural properties and implications
  for dark matter annihilation signals}.
\newblock \emph{Mon. Not. Roy. Astron. Soc.}, 466\penalty0 (4):\penalty0
  4974--4990, 2017.
\newblock \doi{10.1093/mnras/stx026}.

\bibitem[Navarro et~al.(1996)Navarro, Frenk, and White]{Navarro:1995iw}
J.~F. Navarro, C.~S. Frenk, and S.~D.~M. White.
\newblock {The Structure of cold dark matter halos}.
\newblock \emph{Astrophys. J.}, 462:\penalty0 563--575, 1996.
\newblock \doi{10.1086/177173}.

\bibitem[Palacio(2018)]{Palacio:thesis}
J.~Palacio.
\newblock \emph{Indirect dark matter searches on the Triangulum~II dwarf
  spheroidal galaxy and on the Perseus galaxy cluster with the MAGIC
  Telescopes}.
\newblock PhD thesis, Universitat Autonoma de Barcelona, 2018.
\newblock URL \url{http://www.tdx.cat/handle/10803/462764}.

\bibitem[Peebles(1994)]{Peebles:1994xt}
P.~J.~E. Peebles.
\newblock \emph{{Principles of physical cosmology}}.
\newblock 1994.

\bibitem[Pinzke and Pfrommer(2010)]{Pinzke:2010st}
A.~Pinzke and C.~Pfrommer.
\newblock {Simulating the gamma-ray emission from galaxy clusters: a universal
  cosmic ray spectrum and spatial distribution}.
\newblock \emph{Mon. Not. Roy. Astron. Soc.}, 409:\penalty0 449, 2010.
\newblock \doi{10.1111/j.1365-2966.2010.17328.x}.

\bibitem[Pinzke et~al.(2009)Pinzke, Pfrommer, and Bergstrom]{Pinzke:2009cp}
A.~Pinzke, C.~Pfrommer, and L.~Bergstrom.
\newblock {Gamma-rays from dark matter annihilations strongly constrain the
  substructure in halos}.
\newblock \emph{Phys. Rev. Lett.}, 103:\penalty0 181302, 2009.
\newblock \doi{10.1103/PhysRevLett.103.181302}.

\bibitem[Pinzke et~al.(2011)Pinzke, Pfrommer, and Bergstrom]{Pinzke:2011ek}
A.~Pinzke, C.~Pfrommer, and L.~Bergstrom.
\newblock {Prospects of detecting gamma-ray emission from galaxy clusters:
  cosmic rays and dark matter annihilations}.
\newblock \emph{Phys. Rev.}, D84:\penalty0 123509, 2011.
\newblock \doi{10.1103/PhysRevD.84.123509}.

\bibitem[Reiprich and Boehringer(2000)]{Reiprich:1999qm}
T.~H. Reiprich and H.~Boehringer.
\newblock {The mass function of an x-ray flux-limited sample of galaxy
  clusters}.
\newblock \emph{Nucl. Phys. Proc. Suppl.}, 80:\penalty0 0917, 2000.

\bibitem[{Roos}(2010)]{Roos:2010}
M.~{Roos}.
\newblock {Dark Matter: The evidence from astronomy, astrophysics and
  cosmology}.
\newblock \emph{ArXiv e-prints}, Jan. 2010.

\bibitem[S\'anchez-Conde et~al.(2011)S\'anchez-Conde, Cannoni, Zandanel, Gomez,
  and Prada]{SanchezConde:2011ap}
M.~A. S\'anchez-Conde, M.~Cannoni, F.~Zandanel, M.~E. Gomez, and F.~Prada.
\newblock {Dark matter searches with Cherenkov telescopes: nearby dwarf
  galaxies or local galaxy clusters?}
\newblock \emph{JCAP}, 1112:\penalty0 011, 2011.
\newblock \doi{10.1088/1475-7516/2011/12/011}.

\bibitem[Sj{\"o}strand et~al.(2015)Sj{\"o}strand, Ask, Christiansen, Corke,
  Desai, Ilten, Mrenna, Prestel, Rasmussen, and Skands]{Sjostrand:2014zea}
T.~Sj{\"o}strand, S.~Ask, J.~R. Christiansen, R.~Corke, N.~Desai, P.~Ilten,
  S.~Mrenna, S.~Prestel, C.~O. Rasmussen, and P.~Z. Skands.
\newblock {An Introduction to PYTHIA 8.2}.
\newblock \emph{Comput. Phys. Commun.}, 191:\penalty0 159--177, 2015.
\newblock \doi{10.1016/j.cpc.2015.01.024}.

\bibitem[Smith et~al.(2016)]{Smith:2015qhs}
G.~P. Smith et~al.
\newblock {LoCuSS: Testing hydrostatic equilibrium in galaxy clusters}.
\newblock \emph{Mon. Not. Roy. Astron. Soc.}, 456\penalty0 (1):\penalty0
  L74--L78, 2016.
\newblock \doi{10.1093/mnrasl/slv175}.

\bibitem[Sun et~al.(2005)Sun, Jerius, and Jones]{Sun:2005yd}
M.~Sun, D.~Jerius, and C.~Jones.
\newblock {A Small x-ray corona of the narrow-angle tail radio galaxy NGC 1265
  soaring through the Perseus cluster}.
\newblock \emph{Astrophys. J.}, 633:\penalty0 165--173, 2005.
\newblock \doi{10.1086/452620}.

\bibitem[Taylor et~al.(2006)Taylor, Gugliucci, Fabian, Sanders, Gentile, and
  Allen]{Taylor:2006ta}
G.~B. Taylor, N.~E. Gugliucci, A.~C. Fabian, J.~S. Sanders, G.~Gentile, and
  S.~W. Allen.
\newblock {Magnetic fields in the center of the perseus cluster}.
\newblock \emph{Mon. Not. Roy. Astron. Soc.}, 368:\penalty0 1500--1506, 2006.
\newblock \doi{10.1111/j.1365-2966.2006.10244.x}.

\bibitem[Zhao(1996)]{Zhao:1995cp}
H.~Zhao.
\newblock {Analytical models for galactic nuclei}.
\newblock \emph{Mon. Not. Roy. Astron. Soc.}, 278:\penalty0 488--496, 1996.
\newblock \doi{10.1093/mnras/278.2.488}.

\bibitem[Zimmer et~al.(2011)Zimmer, Conrad, and Pinzke]{Zimmer:2011vy}
S.~Zimmer, J.~Conrad, and A.~Pinzke.
\newblock {A Combined Analysis of Clusters of Galaxies - Gamma Ray Emission
  from Cosmic Rays and Dark Matter}.
\newblock 2011.

\end{thebibliography}

\end{document}